\documentclass[11pt,a4paper]{article}
\pdfoutput=1
\usepackage{jheppub,hyperref}
\usepackage{multirow, graphicx,amssymb,url,mathrsfs,amsmath}
\usepackage{wrapfig,boxedminipage,setspace,subfigure,epsfig}
\usepackage{amsxtra,amstext,latexsym,dsfont,amsfonts}
\usepackage{color,eucal}

\renewcommand{\a}{\alpha}

\newcommand{\be}{\begin{equation}}
\newcommand{\ee}{\end{equation}} 
\newcommand{\ba}{\begin{array}}
\newcommand{\ea}{\end{array}}

\title{Interaction induced quasi-particle spectrum and the origin of the pinning peak in holography} 
\author[a]{Geunho Song,}
\author[b]{Yunseok Seo,}
\author[c]{Keun-Young Kim,}
\author[a]{and Sang-Jin Sin}
\emailAdd{sgh8774@gmail.com}
\emailAdd{yseo@gist.ac.kr}
\emailAdd{fortoe@gist.ac.kr}
\emailAdd{sjsin@hanyang.ac.kr}
\affiliation[a]{ Department of Physics, Hanyang University, Seoul 04763, Korea }
\affiliation[b]{ GIST college, Gwangju Institute of Science and Technology, Gwangju 61005, Korea  }
\affiliation[c]{ School of Physics and Chemistry, Gwangju Institute of Science and Technology, Gwangju 61005 Korea
}
 \abstract{  It is often said that  interactions destroy 
 the particle nature of excitations. We report that,  in holographic theory adding interaction term can create a new quasi particle spectrum, on the contrary. We show this by calculating the optical conductivity in a model  with exact background solution 
 and finding a new  quasi-particle spectrum. 
   We argue  that  such new poles are generic consequence of any non-minimal interaction  like  Chern-Simon term. 
The interaction driven metal-insulator  transition and  the pinning effect in holography are examples of this phenomena. 
We also point out that the origin of the pinning peak is the vortex formation by the anomalous magnetic moment induced by the interaction term.
}
\begin{document}
\maketitle

\section{Introduction}
The effects of  an interaction in   weakly interacting   theory are of two folds: one is to smooth out the particle character of the   basic excitations and the other is to mix and deform spectral curves which exist before we add the interaction term.  
On the other hand, for  the Lagrangian with  `free' holographic bulk field,  particle spectrum of boundary  dual theory is already washed out at least partially through  the  minimal coupling of the field to  the  bulk gravity.  
This can be demonstrated  in the bulk  fermion's Green function,  $G_{R}\sim k^{2m-1}  \gamma\cdot k$ \cite{Iqbal:2009fd}.
Then what is the role  of the bulk interaction term?  

\begin{figure}[]
\centering
    \subfigure[Without the Pauli term]
   {\includegraphics[width=6cm]{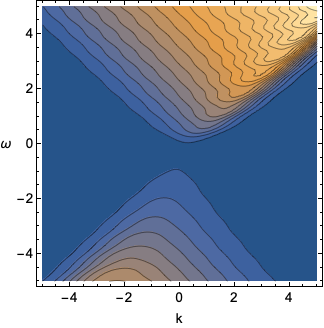} \label{}}
   \hspace{1cm}
      \subfigure[With the Pauli term]
   {\includegraphics[width=6cm]{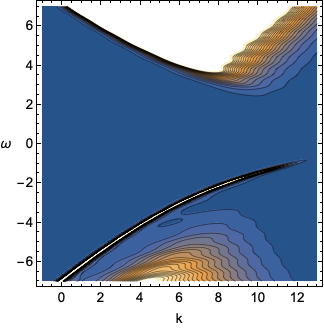} \label{}}
   \caption{ Fermion spectral functions.  (a) Before  the Pauli term is added. No particle nature.   (b) After the Pauli term is added, a new branch appears, implying  new quasi-particle creation.  The background used is the RN black hole with chemical potential $\mu=\sqrt3$, temperature $T=0.1$ and $p=0$ and 10 for (a) and (b) respectively. 
   \label{fig:fermion}}
           \end{figure}

In this paper, we will show that  the bulk interaction term can create a new  sharp peak for the system where original spectrum is almost featureless. This is somewhat opposite phenomena to that of  weakly interacting system. In the  fermion spectral function with bulk mass $m$, there was a hint already:  in search of  the gap creation  using the Pauli     term, $p {\bar \psi} \gamma^{\mu\nu}F_{\mu\nu}\psi$ \cite{Edalati:2010ww}, it was observed \cite{Seo:2018hrc} that a new branch of spectral curve  is created after adding the Pauli term. 
It  was based on the parameter regime of $m\approx -0.5$ where the spectral function mimics  that of the weakly interacting theory.   However,  even for  the regime  $m\approx 0.5$ where particle natures are washed out, a new branch  with a sharp particle feature appears for a large enough $p$, which  is unlikely to happen in weakly interacting systems. See the figure \ref{fig:fermion} (a) and (b).
with and without the Pauli term. 
 
Here,   we want to discuss  the phenomena in the context of transport    so that we can discuss  phenomena like metal-insulator transition more directly. 
For this reason,  we consider  the optical conductivity by considering the bulk   field $A_{\mu}$ which is dual to the current   $J_{\mu}$.  The optical conductivity can also be considered as a measure of spectral function of vector particles in the boundary. We use a background that has analytical expression. This is a model \cite{Seo:2015pug,Seo:2017yux} where a charged black hole is deformed by introducing  an interaction term  describing the magnetic impurities coupled to the `instanton density' $F\wedge F$.  A quasi particle peak develops as we increase the strength of interaction of the magnetic impurity and the peak becomes   sharper for a larger coupling.  This phenomena can be described as  ``bulk  interaction term is the  dual of the boundary quasi-particle''.

The study of  the change of holographic system  induced by the interaction term has some history.  
In \cite{Nakamura:2009tf,Donos:2013gda,Erdmenger:2013zaa,Jokela:2014dba,Ling:2014saa,Donos:2014oha,Cai_2017}, it was noted that adding the Chern-Simon 
term $A\wedge F\wedge F$ can lead to the charge density wave instability. 
In \cite{Donos:2012js},  in a theory with two gauge fields $A_{1},A_{2}$, it was reported that a gap in optical conductivity can be created   in the presence of  a Chern-Simon like term $A_{2}\wedge F_{2}\wedge F_{1}$ if $A_{2}$ is magnetically sourced and the physics was identified as metal-insulator transition. 

More recently, in \cite{Jokela_2017, Andrade_2018, Donos:2019hpp, SciPostPhys.3.3.025},  a shifted peak in optical conductivity was interpreted  as the pinning effect \cite{PhysRevB.17.535,PhysRevLett.45.935,Fukuyama:1978ab}.   Namely,  if we consider the acoustic phonon as a Goldstone boson of translational symmetry breaking,  
 one asks what is the fate of it  when we add  small explicit symmetry breaking, they suggest that  it disappears by getting mass.  In these context, we want to answer how it gets the mass in more   explicit way. 
 We will suggest that the origin of the pinning pole in general by comparing it with the cyclotron peak and identify it as    the vortex formation by the anomalous magnetic moment induced by the interaction term. The metal insulator transition as well as the holographic pinning  can be viewed in the  framework of spectral reconfiguration.  
 
The rest of the paper goes as follows. 
In section 2, after   describing  the methods   we will calculate  the AC conductivity as a spectral measure   to demonstrate the appearance of new quasi-particle and we discuss the origin of the new quasi particle peak.  We conclude with discussions at section 4. 
   
 \section{Setup and  methods}
In this section we describe the basic set up and calculational methods for optical conductivity. We adopted  the basic scheme of \cite{Kim:2014bza} for our purpose. 
The readers who are not interested in technical details can go to the section 3 directly.  
\subsection{Action and background geometry}
We start from the Einstein-Maxwell action with two massless scalars, one of which is coupled with the gauge instanton density on 4-d manifold $\mathcal{M}$ with boundary $\partial \mathcal{M}$.
\begin{align}\label{eq:action}
	2\kappa^2S_0&=\int_{\mathcal{M}} d^4x \sqrt{-g}\left\{ R+\frac{6}{L^2}-\frac{1}{4}F^2-\sum_{I,a=1,2}\frac{1}{2}(\partial\chi_I^{(a)})^2\right\} +\\
	&\qquad\quad -\frac{q_{\chi}}{16}\int_{\mathcal{M}}(\partial\chi_I^{(2)})^2\epsilon^{\mu \nu\rho\sigma}F_{\mu \nu} F_{\rho\sigma}-2\int_{\partial\mathcal{M}}d^3x\sqrt{-\gamma}K,\\
	S_c&=-\int_{\partial\mathcal{M}}d^3x\sqrt{-\gamma}\left(\frac{4}{L}+R[\gamma]-\sum_{I,a=1,2}\frac{L}{2}\nabla\chi_I^{(a)}\cdot\nabla\chi_I^{(a)}\right),
\end{align}
where $q_{\chi}$ is a coupling, and $\kappa^2=8\pi G$ and $L$ is the AdS radius and we set $2\kappa^2=L=1.$ $S_c$ is the counter term for holographic renormalization. The finite renormalized on-shell action is
\begin{align}
	\label{renaction}S_{ren}=\lim_{\Lambda\rightarrow\infty}(S_0+S_c)_{\text{on-shell}}.
\end{align}
The corresponding equations of motion are given by
\begin{align}\label{eom1}
&\partial_{\mu}\left(\sqrt{-g} g^{\mu\nu}\sum_I\left( \partial_{\nu} \chi_{I}^{(a)} \right)\right) +\frac{q_{\chi}}{8}\, \partial_{\mu} \left( \epsilon^{\rho\sigma\lambda\gamma} F_{\rho\sigma} F_{\lambda\gamma} g^{\mu\nu}\sum_{I,a}\left(\partial_{\nu} \chi_{I}^{(a)}\delta_{2a}\right)\right)=0, \\
\label{eom2}&\partial_{\mu} \left( \sqrt{-g} F^{\mu\nu} +\frac{q_{\chi}}{4} g^{\rho\sigma} \sum_{I,a}\left( \partial_{\rho} \chi_{I}^{(a)} \,\partial_{\sigma} \chi_{I}^{(a)}\delta_{2a}\right) \epsilon^{\alpha\beta \mu \nu}F_{\alpha\beta}\right )=0, \\
\label{eom3}&R_{\mu\nu}-\frac{1}{2} g_{\mu\nu}\left[ R+6-\frac{1}{4}F^2-\sum_{I,a}\frac{1}{2}(\partial\chi_I^{(a)})^2 \right] -\frac{1}{2} F^{\rho}_{\mu} F_{\rho\nu}-\frac{1}{2}\sum_{I,a}(\partial_{\mu}\chi_{I}^{(a)})(\partial_{\nu}\chi_{I}^{(a)})\cr
&-\frac{1}{\sqrt{-g}}\, \frac{q_{\chi}}{16} \sum_{I,a} (\partial_{\mu} \chi_{I}^{(a)})(\partial_{\nu}\chi_{I}^{(a)})\delta_{2a} \epsilon^{\rho\sigma\lambda\gamma} F_{\rho\sigma}F_{\lambda \gamma} =0.
\end{align}
With the following ansatz, 
 \begin{align}\label{bgsol0}
 	&A=a(r)dt+\frac{1}{2}H(xdy-ydx),\nonumber\\
 	&\chi_I^{(1)}=\left( \begin{array}{c}
							\alpha x  \\
							\alpha y
\end{array} \right),
\qquad 
	\chi_I^{(2)}=\left( \begin{array}{c}
							\beta x  \\
							\beta y
\end{array} \right)\nonumber\\
 	&ds^2=-U(r)dt^2+\frac{dr^2}{U(r)}+r^2(dx^2+dy^2),
 \end{align}
 we  found the exact solutions as follows:
 \begin{align}\label{bgsol}
 	&U(r)=r^2-\frac{\alpha^2+\beta^2}{2}-\frac{m_0}{r}+\frac{q^2+H^2}{4r^2}+\frac{\beta^4H^2q_{\chi}^2}{20r^6}-\frac{\beta^2Hqq_{\chi}}{6r^4},\\
 	&a(r)=\mu-\frac{q}{r}+\frac{\beta^2Hq_{\chi}}{3r^3},
 \end{align}
 where $\mu$ is the chemical potential, $q$ and $m_0$ are determined by the conditions $A_t(r_0)=U(r_0)=0$ at the black hole horizon $(r=r_0)$. $q$ is the conserved $U(1)$ charge density and $\alpha$, $\beta$ correspond to the non-magnetic impurity  and magnetic impurity density respectively.
 \begin{align}\label{m0q}
 	q&=\mu r_0 +\frac{1}{3}\theta H \qquad \text{with} \quad \theta=\frac{\beta^2q_{\chi}}{r_0^2}\\
 	m_0&=r_0^3\left(1+\frac{r_0^2\mu^2+H^2}{4r_0^4}-\frac{\alpha^2+\beta^2}{2r_0^2}\right)+\frac{\theta^2H^2}{45r_0}
 \end{align}
The temperature of the boundary system can be derived from the Hawking temperature in the bulk,
\begin{align}\label{temperature}
4\pi T = U'(r_0)=3 r_0 -\frac{1}{4 r_0^3}\left[ H^2 +2 r_0^2 (\alpha^2+\beta^2) +(q -H \theta)^2   \right] \,,
\end{align}

\subsection{Linearized equations of motion and the on-shell action}
To compute the optical conductivity, we need to consider small perturbations around the background (\ref{bgsol0}) and (\ref{bgsol}) to use linear response theory. 
 We turn on small fluctuations $\delta g_{ti}, \delta A_i,$ and $\delta \chi_i$  with $i=x,y$, around the  background at zero momentum $(\vec{k}=0)$. The fields $\delta g_{ti}$ and $\delta A_i$ are related  to the temperature gradient and the electric field respectively.   Now  we add $\delta \chi_i$ for momentum relaxation for finite conductivity.   For optical conductivities, we should consider a gauge $\delta g_{ri}=0$, and this gauge does not change any physical results. Indeed, the zero frequency limit of optical conductivity in this gauge ($\delta g_{ri}=0$) is consistent with the DC conductivity which is analytically obtained with non-trivial gauge of $\delta g_{ri}$.
\\These fields, $\delta g_{ti}, \delta A_i$ and $\delta \chi_i^{(a)}$, can be expressed in momentum space as
\begin{align}
	\delta A_i&=\int_{-\infty}^{\infty}\frac{d\omega}{2\pi} e^{-i\omega t}a_i(r),\\
	\delta g_{ti}&=\int_{-\infty}^{\infty}\frac{d\omega}{2\pi} e^{-i\omega t}r^2h_{ti}(r)\label{metflu}\\
	\delta \chi_i^{(1)}&=\int_{-\infty}^{\infty}\frac{d\omega}{2\pi} e^{-i\omega t}\phi_i(r)\\
	\delta \chi_i^{(2)}&=\int_{-\infty}^{\infty}\frac{d\omega}{2\pi} e^{-i\omega t}\psi_i(r),
\end{align}
where $h_{ti}$ goes to constant at the boundary $(r\rightarrow \infty)$ because of $r^2$ in the metric fluctuations (\ref{metflu}) and we set $r_0=1$ for convenience of numerical analysis. From the equations of motion (\ref{eom1})-(\ref{eom3}), the linearised equations in momentum space are given by the following 8 equations.\\
- Einstein equations:
\begin{align}
	\label{em1} 0=h_{ti}''&+\frac{4}{r} h_{ti}'+\frac{(q-\theta H/r^2)}{r^4}a_x'-\frac{H^2}{r^4U}h_{ti}-\frac{i\alpha\omega}{r^2U}\phi_i \cr
	&-\frac{(\beta^2r^4-q\theta H-\theta^2 H^2/r^2)(\beta h_{ti}+i\omega\psi_i)}{\beta r^6U}-\frac{\alpha^2}{r^2U}h_{ti}-\epsilon_{ij}\frac{i\omega H}{r^4U} a_j 	\\
0=	\label{em2}h_{ti}'&+\epsilon_{ij}\frac{iUH}{r^4\omega}a_j'+\frac{i\beta U(r^6-q_{\chi}qHr^2+q_{\chi}^2H^2\beta^2)}{r^8\omega}\psi_i'+\frac{i\alpha U}{r^2\omega}\phi'_i \cr
	&+\frac{(qr^2-q_{\chi}H\beta^2)(\omega a_x+iH\epsilon_{ij}h_{tj})}{r^6\omega} 
\end{align}
- Maxwell equations:
\begin{align}\label{max1}
	0=a_i'' &+\frac{U'}{U}a_i'+\frac{(q-\theta H/r^2)}{U}h_{ti}'+\frac{\omega^2}{U^2}a_i +\epsilon_{ij}\left(\frac{2i\theta\omega}{r^3U}a_j+\frac{iH\omega}{U^2}h_{tj}\right) 
	+\frac{2\theta H }{r^3U}h_{ti}
\end{align}
- Scalar equations:
\begin{align}
\label{sc2}
0&=\phi_i''+\left(\frac{2}{r}+\frac{U'}{U}\right)\phi_i'-\frac{i\alpha\omega}{U^2} h_{tx}+\frac{\omega^2}{U}\phi_i \\
\label{sc1}0&=\left(1-\frac{q\theta H}{\beta^2r^4}+\frac{\theta^2 H^2}{\beta^2r^6}\right)\psi_i''+\left(\frac{2}{r}+\frac{2q\theta H}{\beta^2r^5}-\frac{4H^2\theta^2}{\beta^2r^7}+\frac{U'}{U}-\frac{q\theta HU'}{\beta^2r^4U}+\frac{H^2\theta^2 U'}{\beta^2r^6U}\right) \psi_i'  \cr
 &\qquad   +\frac{\omega(\beta^2 r^4-qH\theta-\theta^2 H^2/r^2)(\omega \psi_i-i\beta h_{ti})}{\beta^2r^4U^2}
\end{align} 
Among these 10 equations, only 6 are independent. Indeed, differentiating (\ref{em2}) with respect to $r$ and putting background solution into the equation, then we can obtain (\ref{em1}). We need to find the boundary conditions to solve these equations. There are two boundary conditions, which are the incoming boundary conditions at the horizon and the Dirichlet boundary conditions at the boundary.
Near the black hole horizon $(r\rightarrow 1)$ the solutions are expanded as
 \begin{align}\label{horizon}
 	h_{ti}&=(r-1)^{\nu_{\pm}+1}(h_{ti}^{(I)}+h_{ti}^{(II)}(r-1)+\cdots),\nonumber\\
 	a_i&=(r-1)^{\nu_{\pm}}(a_i^{(I)}+a_i^{(II)}(r-1)+\cdots),\nonumber\\
	\phi_i&=(r-1)^{\nu_{\pm}}(\phi_i^{(I)}+\phi_i^{(II)}(r-1)+\cdots)\nonumber\\
 	\psi_i&=(r-1)^{\nu_{\pm}}(\psi_i^{(I)}+\psi_i^{(II)}(r-1)+\cdots)
 \end{align}
where 
\be
\nu_{\pm}=\pm 36i\omega/(H^2(9+4q_{\chi}^2\beta^4)-12q_{\chi}H\beta^2\mu+9(-12+2\beta^2+\mu^2)),
\ee and the incoming boundary condition corresponds to taking $\nu_+$ and discarding $\nu_{-}$. It turns out that only  6 coefficients, $ a_i^{(I)}, \psi_i^{(I)}$ and $\phi_i^{(I)}$, of horizon behavior of the solutions can be chosen independently because $h_{ti}^{(I)}$ and all other higher order coefficients can be determined by them.\\
 \indent The asymptotic solutions near the boundary $(r\rightarrow\infty)$ can be written by
 \begin{align}\label{boundary}
 	h_{ti}&=h_{ti}^{(0)}+\frac{1}{r^2}h_{ti}^{(2)}+\frac{1}{r^3}h_{ti}^{(3)}+\cdots,\nonumber\\
 	a_i&=a_i^{(0)}+\frac{1}{r}a_i^{(1)}+\cdots,\nonumber\\
	\phi_i&=\phi_i^{(0)}+\frac{1}{r^2}\phi_i^{(2)}+\frac{1}{r^3}\phi_i^{(3)}+\cdots,\nonumber\\
 	\psi_i&=\psi_i^{(0)}+\frac{1}{r^2}\psi_i^{(2)}+\frac{1}{r^3}\psi_i^{(3)}+\cdots.
  \end{align}
\indent With the incoming boundary condition and chosen free parameters $a_i^{(I)}, \phi_i^{(I)},$ and $\psi_i^{(I)}$, we can numerically determine the leading and subleading terms by solving linearised equations of motion (\ref{em1})-(\ref{sc1}). The leading terms play the role of sources for the operators whose expectation values are encoded in $h_{ti}^{(3)}, a_i^{(1)},\phi_i^{(3)},$ and $\psi_i^{(3)}$ respectively.\\
 We expand the renormalized action (\ref{renaction}) around the modified RN-$AdS_4$ background (\ref{bgsol0})-(\ref{bgsol}) and using the equations of motion to obtain a quadratic on-shell action:
 \begin{align}
 	 S_{ren}^{(2)}&=\lim_{\Lambda \to \infty}\frac{1}{2}\int_{r=\Lambda}d^3x\Bigg{[}\delta \tilde{h}_{ti}\left(\left(-q+\frac{\theta H}{r^2}\right)\delta A_i+\frac{(\alpha\delta\dot{\chi}_i^{(1)}+\beta\delta\dot{\chi}_i^{(2)}) r^2}{\sqrt{U(r)}} \right)\hfill \nonumber\\ 
& +\left(4r^3-\frac{4r^4}{\sqrt{U(r)}}\right)\delta \tilde{h}_{ti}^2-r^2\delta\chi_i^{(1)}\delta\chi_i^{(1)\prime}-\frac{(\beta^2r^4-q\theta H+\theta^2H^2/r^2)U(r)}{\beta^2 r^2}\delta\chi_i^{(2)}\delta\chi_i^{(2)\prime}\nonumber\\
 &	+\epsilon_{ij}\frac{(q\theta H-\theta^2H/r^2)U(r)}{\beta^2 r^2}\delta\chi_i^{(2)}\delta\chi_j^{(2)\prime} +\epsilon_{ij}\frac{\theta}{2r^2}\delta \dot{A}_i\delta A_j-U(r)\delta A_i\delta A_i' \nonumber\\
&-\frac{\alpha r^2}{\sqrt{U(r)}}\delta\chi_i^{(1)}\tilde{h}_{ti}+\frac{r^2}{\sqrt{U(r)}}\delta\chi_i^{(1)}\delta\ddot{\chi}_i^{(1)}
-\frac{\beta r^2}{\sqrt{U(r)}}\delta\chi_i^{(2)}\tilde{h}_{ti}+\frac{r^2}{\sqrt{U(r)}}\delta\chi_i^{(2)}\delta\ddot{\chi}_i^{(2)}\Bigg{]}
 \end{align}
where $\delta\tilde{h}_{ti}=r^{-2}\delta g_{ti}(t,r)$, dot denotes time derivative, prime denotes $r$-derivative. With the expression of fluctuation in momentum space, we can rewrite the quadratic action  in momentum space:
\begin{align}\label{quadaction1}
	S_{ren}^{(2)}=&\frac{\mathcal{V}_2}{2}\int_0^{\infty}\frac{d\omega}{2\pi}\left(-q\bar{a}_i^{(0)}h_{ti}^{(0)}-2m_0\bar{h}_{ti}^{(0)}h_{ti}^{(0)}+\bar{a}_i^{(0)}a_i^{(1)}-3\bar{h}_{ti}^{(0)}h_{ti}^{(3)}+\bar{\phi}_i^{(0)}\phi_i^{(3)}+\bar{\psi}_i^{(0)}\psi_i^{(3)}\right)\nonumber \\
	&\qquad \qquad\qquad+h.c,
\end{align}
where $\mathcal{V}_2$ is two dimensional spatial volume $\int dxdy$. It is understood that the argument of the variables is $\omega$ while that of those with the bar is $-\omega$. 

\subsection{Optical conductivities}
We introduce a systematic numerical method with multi-fields and constraints for our case briefly\cite{Kim:2014bza}. Let us start with $n$ fields $\Phi^a(t,r),$ $(a=1,2,\cdots,n)$, which are fluctuations around a background. Assume that they satisfy a set of coupled $n$ second order differential equations and all the fluctuation fields depend only on $t$ and $r$:
\begin{align}
	\Phi^a(t,r)=\int\frac{d\omega}{2\pi}e^{-i\omega t}r^p\Phi_{\omega}^a(r),
\end{align}
where $r^p$ is multiplied such that the solution $\Phi_{\omega}^a(r)$ has constant behavior at the boundary $(r\rightarrow\infty)$. For example, $p=2$ in (\ref{metflu}).\\
 \indent Near the black hole horizon $(r=1)$, solutions are expanded as
 \begin{align}
 	\Phi^a(r)=(r-1)^{\nu_{a\pm}}(\phi^a+\tilde{\phi}^a(r-1)+\cdots),
 \end{align}
where we dropped the subscript $\omega$ for simplicity and $\pm$ denotes incoming/outgoing boundary conditions respectively. For computing the retarded Green's function holographically, we should choose the incoming boundary condition. This choice eliminates the half of independent parameters from $2n$ to $n$. And if we have $n_c$ constraint equations, then the number of independent parameters can be reducted by $n_c$. Consequently, the number of independent parameter is $n-n_c$ so we must choose $n-n_c$ initial conditions, denoted by $\phi^a_{\hat{i}}$ $(\hat{i}=1,2,\cdots,n-n_c)$:
\begin{align}
\left(\phi^a_1\ \phi^a_2\ \phi^a_3\ \cdots\ \phi^a_{n-n_c}\right)=
\left(\begin{array}{cccc}
1&1&1&\cdots\\
1&-1&1&\cdots\\
1&1&-1&\cdots\\
\vdots&\vdots&\vdots&\ddots\\
1&1&1&\cdots
\end{array}\right).
\end{align}
Each set of initial condition $\phi^a_{\hat{i}}$ generates a solution with the incoming boundary condition, denoted by $\Phi^a_{\hat{i}}(r)$, which can be expanded near the boundary as following:
\begin{align}
	\Phi^a_{\hat{i}}(r)\rightarrow \mathbb{S}_{\hat{i}}^a+\cdots+\frac{\mathbb{O}_{\hat{i}}^a}{r^{\delta_a}}+\cdots \qquad\text{(near boundary)},
\end{align}
where $\mathbb{S}_{\hat{i}}^a$ correspond to the sources, which are the leading terms of $\hat{i}$-th solution, and $\mathbb{O}_{\hat{i}}^a$ are relavent to the expectation values of the operators corresponding to the sources $\mathbb{S}_{\hat{i}}^a$ $(\delta_a\geq 1)$.\\
 \indent The general solution can be obtained as a linear combination of them. Let
 \begin{equation}
 	\Phi^a(r)\equiv\Phi_{\hat{i}}^a(r)c^{\hat{i}}\rightarrow \mathbb{S}_{\hat{i}}^ac^{\hat{i}}+\cdots+\frac{\mathbb{O}_{\hat{i}}^ac^{\hat{i}}}{r^{\delta_a}}+\cdots \qquad \text{(near boundary)},
 \end{equation}
with real constants $c^i$'s. We want to choose $c^{\hat{i}}$ such that the leading term in the general solution ($i.e$ $\Phi_{\hat{i}}^a(r)c^{\hat{i}}$) is identified with the independent sources $J^a$.  But if there are constraints, we need the additiononal set of solutions since $a>\hat{i}$. In this case, however, we can generate $n_c$ solutions by using residual gauge transformations.\\
 \indent In our case, $n=8$ and $n_c=2$, which corresponds to the constraints $g_{ri}=0$. There are two sets of additional constant solutions of the equations of motion (\ref{em1})-(\ref{sc1})
 \begin{equation}
 	h_{ti}=h_{ti}^0,\qquad a_i=-\frac{iH\epsilon_{ij}h_{tj}^0}{\omega},\qquad \phi_i=\frac{i\alpha h_{ti}^0}{\omega},\qquad \psi_i=\frac{i\beta h_{ti}^0}{\omega}
 \end{equation}
 where $h_{ti}^0$ is arbitrary constant and $i,j=x,y$. Therefore, the explicit expression for $\mathbb{S}_{\bar{i}}^a$ is 
 \begin{align}
	\mathbb{S}_{\hat{i}}^a=\left(\mathbb{S}_7^a\ \mathbb{S}_8^a\right)=\left(
	\begin{array}{cc}
	0&-i\frac{H}{\omega}\\
	i\frac{H}{\omega}&0\\
	1&0\\
	0&1\\
	i\frac{\alpha}{\omega}&0\\
	0&i\frac{\alpha}{\omega}\\
	i\frac{\beta}{\omega}&0\\
	0&i\frac{\beta}{\omega}
	\end{array}
	\right).
 \end{align}
 \indent These can be understood as residual gauge transformations corresponding $g_{ri}=0$, which is generated by the vector field $\xi^{\mu}$ whose non-vanishing component is $\xi^i=\epsilon^ie^{-i\omega t}$ with constant $\epsilon^i$. So we add a vector along the residual gauge orbit. Since $\mathcal{L}_{\xi}g_{ti}=-i\omega r^2\xi^i$ and $\mathcal{L}_{\xi}\phi=\beta\xi^i,\ \mathcal{L}_{\xi}A_i=-H\epsilon_{ij}\xi^j$. Hence the most general solution becomes
 \begin{equation}
 	\Phi^a(r)=J^a+\cdots+\frac{R^a}{r^{\delta_a}}+\cdots,
 \end{equation}
 where we defined $J^a$ and $R^a$ as a source and  response. For arbitrary sources $J^a$ we can always find $c^I$
 \begin{equation}
 	c^I=(\mathbb{S}^{-1})^I_aJ^a
 \end{equation}
where $I=1,\ldots,n$. The corresponding response $R^a$ is expressed as
 \begin{equation}\label{responses}
 	R^a=\mathbb{O}^a_Ic^I=\mathbb{O}^a_I(\mathbb{S}^{-1})^I_bJ^b.
 \end{equation}
\indent The general on-shell quadratic action in terms of the sources and the responses can be written as
  \begin{equation}\label{quadaction2}
  	S_{ren}^{(2)}=\frac{\mathcal{V}_2}{2}\int_0^{\infty}\frac{d\omega}{2\pi}[\bar{J}^a\mathbb{A}_{ab}J^b+\bar{J}^a\mathbb{B}_{ab}R^b],
  \end{equation}
where $\mathbb{A}$ and $\mathbb{B}$ are regular matrices of order $n$.
 For example, we can rewrite the action (\ref{quadaction1}) with:
 \begin{align}
 	J^a=\left(\begin{array}{c} a_x^{(0)}\\a_y^{(0)}\\h_{tx}^{(0)}\\h_{ty}^{(0)}\\\phi_x^{(0)}\\ \phi_y^{(0)}\\\psi_x^{(0)}\\ \psi_y^{(0)}\end{array}\right),\quad 
 	R^a=\left(\begin{array}{c} a_x^{(1)}\\a_y^{(1)}\\h_{tx}^{(3)}\\h_{ty}^{(3)}\\\phi_x^{(3)}\\ \phi_y^{(3)}\\\psi_x^{(3)}\\ \psi_y^{(3)}\end{array}\right),\quad 
 	\mathbb{A}=\left(\begin{array}{cccc}0&-q&0&0\\ 0&-2m_0&0&0\\0&0&0&0\\0&0&0&0 \end{array}\right)\otimes \boldsymbol{1}_2,\quad 
 	\mathbb{B}=\left(\begin{array}{cccc}1&0&0&0\\ 0&-3&0&0\\0&0&3&0\\0&0&0&3 \end{array}\right)\otimes \boldsymbol{1}_2 ,
 \end{align}
where $\boldsymbol{1}_2$ is the $2\times 2$ unit matrix. Plugging the relation (\ref{responses}) into the action (\ref{quadaction2}) we have
 \begin{equation}
 	S_{ren}^{(2)}=\frac{\mathcal{V}_2}{2}\int_0^{\infty}\frac{d\omega}{2\pi}\bar{J}^a[			\mathbb{A}_{ab}+\mathbb{B}_{ab}\mathbb{O}^c_I(\mathbb{S}^{-1})^I_b]J^b.
 \end{equation}
\indent To sum up, we need four $n\times n$ matrices, $\mathbb{A},\mathbb{B},\mathbb{S}$ and $\mathbb{O}$ to compute the retarded Green's function. Each component of $\mathbb{A}$ and $\mathbb{B}$ can be read off from the quadratic on-shell action (\ref{quadaction1}). The matrices $\mathbb{S}$ and $\mathbb{O}$ can be obtained by solving a set of the differential equations numerically. Notice that the choice of initial conditions is not relevant to the Green  functions.

    In our model, we can get a $8\times 8$ matrix of the retarded Green functions.   Here, however, we will focus only on the $4\times 4$ sub-matrix corresponding to $a_i^{(0)}$ and $h_{ti}^{(0)}$ in (\ref{boundary}),
  \begin{align}
  \left(\begin{array}{cc}
  G_{JJ}^{ij}&G_{JT}^{ij}\\
  G_{TJ}^{ij}&G_{TT}^{ij}
  \end{array}\right)
  \end{align}
where every $G_{AB}^{ij}$ is a $2\times 2$ retarded Green's function with $i=x,y$ for given $A=J,T$. The lower indices $A,B$ denote the operators corresponding to the sources. i.e. the $a_i^{(0)}$ is a source of the electric current $J^i$ and the $h_{ti}^{(0)}$ is a source of the energy-momentum tensor $T^{ti}$. From the linear response theory, it is well known that response functions and the sources have the following relation:
\begin{align}
	\left(\begin{array}{c}\langle J^i\rangle \\ \langle T^{ti}\rangle\end{array}\right)
	=\left(\begin{array}{cc}
  G_{JJ}^{ij}&G_{JT}^{ij}\\
  G_{TJ}^{ij}&G_{TT}^{ij}
  \end{array}\right)
  \left(\begin{array}{c}a_j^{(0)}\\h_{tj}^{(0)}\end{array}\right).
\end{align}
where $\langle J^i\rangle$, $\langle T^{ti}\rangle$, $a_j^{(0)}$ and $h_{tj}^{(0)}$ are $2\times 1$ column matrices, with $i=x,y$. From this matrix relation, we want to obtain the electric $(\hat{\sigma})$, thermal $(\hat{\bar{\kappa}})$, and thermoelectric $(\hat{\alpha},\hat{\bar{\alpha}})$ conductivities defined as
\begin{align}\label{condmat}
	\left(\begin{array}{c}\langle J^i\rangle \\ \langle Q^{ti}\rangle\end{array}\right)
	=\left(\begin{array}{cc}
 \hat{\sigma}^{ij}&\hat{\alpha}^{ij}T\\
  \hat{\bar{\alpha}}^{ij}T&\hat{\bar{\kappa}}^{ij}T
  \end{array}\right)
  \left(\begin{array}{c}E_j\\-(\nabla_jT)/T\end{array}\right).
\end{align}
where $E_i$ is an electric field, $\nabla_iT$ is a temperature gradient along the $i$ direction and $Q^i$ is a heat current, which is defined by $Q^i=T^{ti}-\mu J^i$. Notice that the electric and heat current here contain the contribution of magnetization which should be subtracted, so we use the conductivities with hat. By taking into account diffeomorphism invariance, (\ref{condmat}) can be expressed as
\begin{align}
\begin{pmatrix}
 \sigma^{ij}&\alpha^{ij}T\\
  \bar{\alpha}^{ij}T&\bar{\kappa}^{ij}T
  \end{pmatrix}
	=\begin{pmatrix}
  -\frac{iG_{JJ}^{ij}}{\omega}&\frac{i(\mu G_{JJ}^{ij}-G_{JT}^{ij})}{\omega}\\
  \frac{i(\mu G_{JJ}^{ij}-G_{JT}^{ij})}{\omega}&-\frac{i(G_{TT}^{ij}-G_{TT}^{ij}(\omega=0)-\mu(G_{JT}^{ij}+G_{TJ}^{ij}-\mu G_{JJ}^{ij}))}{\omega}
  \end{pmatrix}
  -\frac{H}{T}\begin{pmatrix}
  0&\Sigma_1\epsilon^{ij}\\
  \Sigma_1\epsilon^{ij}&\Sigma_2\epsilon^{ij}
  \end{pmatrix}
\end{align}

\section{Interaction  induced quasi-particles in  AC conductivity}
In this section, we investigate magnetic impurity effect on the optical conductivity. To see the effect of the magnetic impurity, we turn off the non-magnetic impurity density parameter, $\alpha=0$ in all calculations.
\subsection{Appearance of a new quasi particle at $\mu = H =0$ }
As the first step to study   $q_{\chi}$ coupling  effect, we  consider the  simplest case where   neither chemical potential nor   external magnetic field exists. Here, the background geometry is  simply the AdS Schwartzschild black hole with momentum relaxation.   
 Then the value of $q_{\chi}$ term itself vanishes  in the background solution,  though its effect should appear in the fluctuation level as one can see in  the Maxwell equation (\ref{max1}). 

\begin{figure}[]
\centering
    \subfigure[ ]
   {\includegraphics[width=6cm]{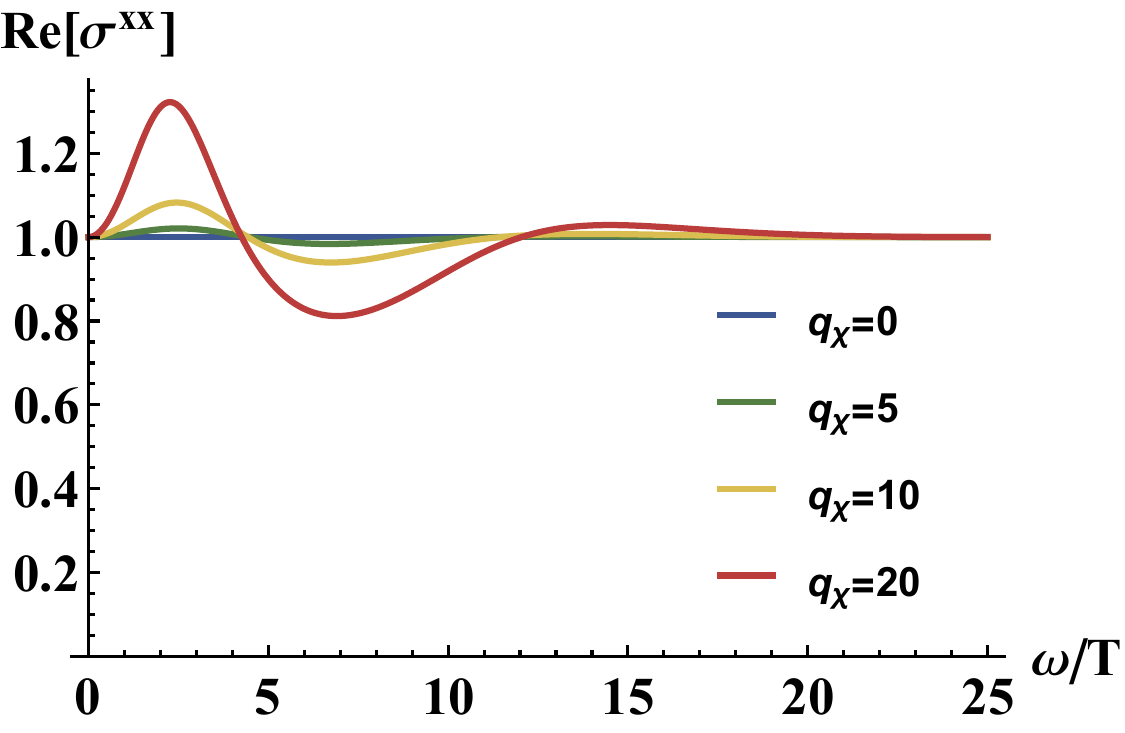} \label{}}
   \hspace{1cm}
       \subfigure[]
   {\includegraphics[width=6cm]{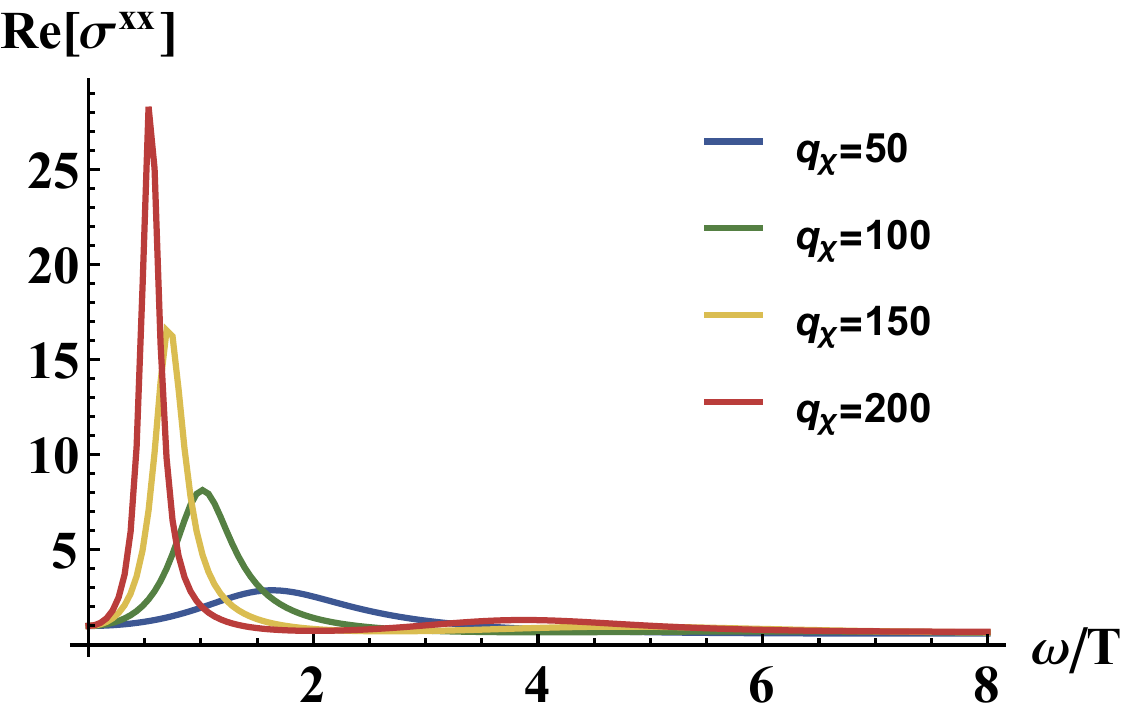} \label{}}
   \caption{ $q_{\chi}$ dependence of optical conductivity for $\mu =0$ and $H =0$. Stronger interaction gives sharper peak at smaller frequency and sum rule is checked. 
           } \label{fig:Sxx01}
           \end{figure}

Figure \ref{fig:Sxx01} shows $q_{\chi}$ dependence of the optical conductivity. 
           \begin{figure}[]
\centering
    \subfigure[ ]
   {\includegraphics[width=4.5cm]{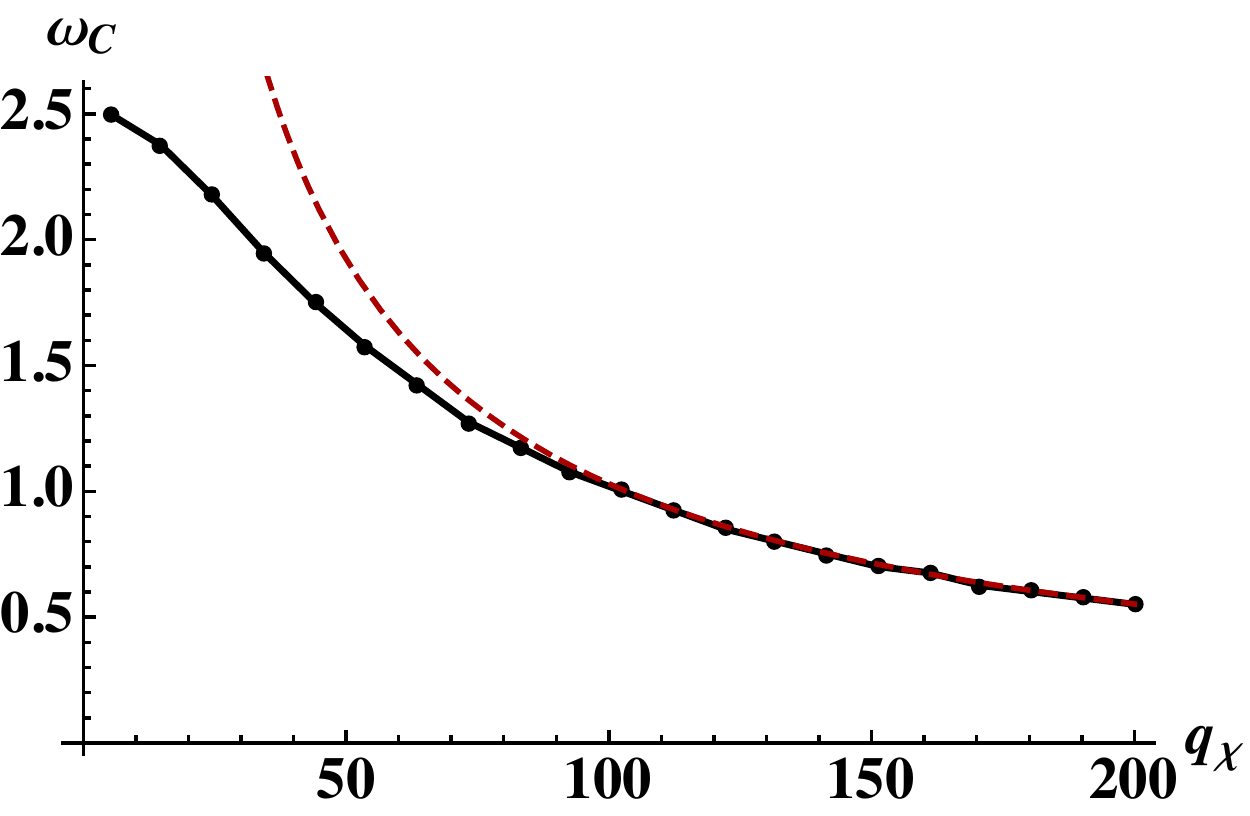} \label{}}
   \hspace{1cm}
       \subfigure[]
   {\includegraphics[width=4.5cm]{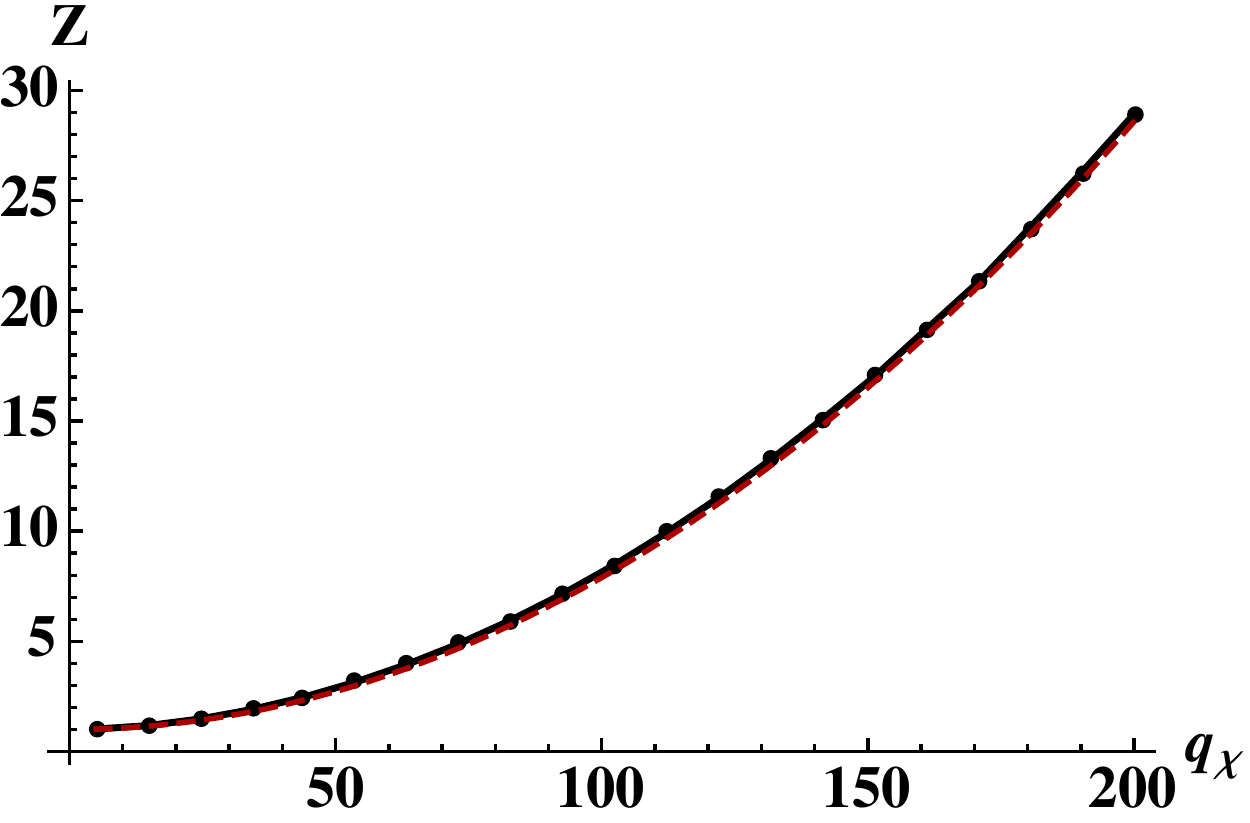} \label{}}
   \caption{ $q_{\chi}$ dependence of  $\omega_C$ (peak frequency) and $Z$ (height of the peak). Dotted lines are $\omega_C = a \,q_{\chi}^{-1}$ (a)  and $Z = b\, q_{\chi}^2 +1$ (b).            } \label{fig:wC}
           \end{figure}
When $q_{\chi}=0$, the optical conductivity is constant,  $\sigma(\omega)=1$, which is attributed to the    electron-hole pair creation. When we turn on $q_{\chi}$, a quasi-particle peak appears as we can see from Figure \ref{fig:Sxx01} (a)(b). 
 As we increase $q_{\chi}$, the peak becomes higher  while the DC conductivity is kept to be one. See figure \ref{fig:Sxx01}(a).   We checked that the whole optical conductivity satisfies a sum rule, so that the total area below the optical conductivity curve is fixed. It means that  the new quasi-particle degree of freedom is created at the price of reducing higher frequency modes.
As we increase the value of $q_{\chi}$ further, the quasi-particle peak becomes   sharper and the peak frequency ($\omega_C$) is decreasing.  See Figure \ref{fig:Sxx01} (b).

From the intuition of weakly interacting theory, this is surprising since one  expects  that interactions smooth out the particle character rather than create it. 
This  feature of the optical conductivity  is   similar to the pinning effect~\cite{Jokela_2017,Andrade_2018, Baggioli_2015,Alberte_2018,Ammon:2019wci, Li_2019,Amoretti:2018tzw, Donos_2019} where charge density wave (CDW) is pinned by impurity so that only   in a window of frequencies the charge density wave (CDW) can slide.  
Therefore, we will call this peak as {\it `pinning peak'}. However, in our case the Drude peak does not disappear 
 and just a new peak is created and therefore a metal-insulator transition does not follow.  

 The excitation frequency  decreases as the interaction strength of impurity increases and  the height of the peak ($Z$) increases such  that 
 \be 
 \omega_C =a q_{\chi}^{-1}, \quad    Z = b q_{\chi}^2 +1,
  \ee
with   some numerical constants $a$ and $b$.
The $q_{\chi}$ dependence of $\omega_C$ and the height of the peak $Z$ are drawn in Figure \ref{fig:wC}.

\subsection{Pinning peak vs. Cyclotron peak: $\mu=0$, $H \ne 0$ }  
\begin{figure}[]
	\centering
\subfigure[Pinning and Cyclotron Peak]
	{\includegraphics[width=4.8cm]{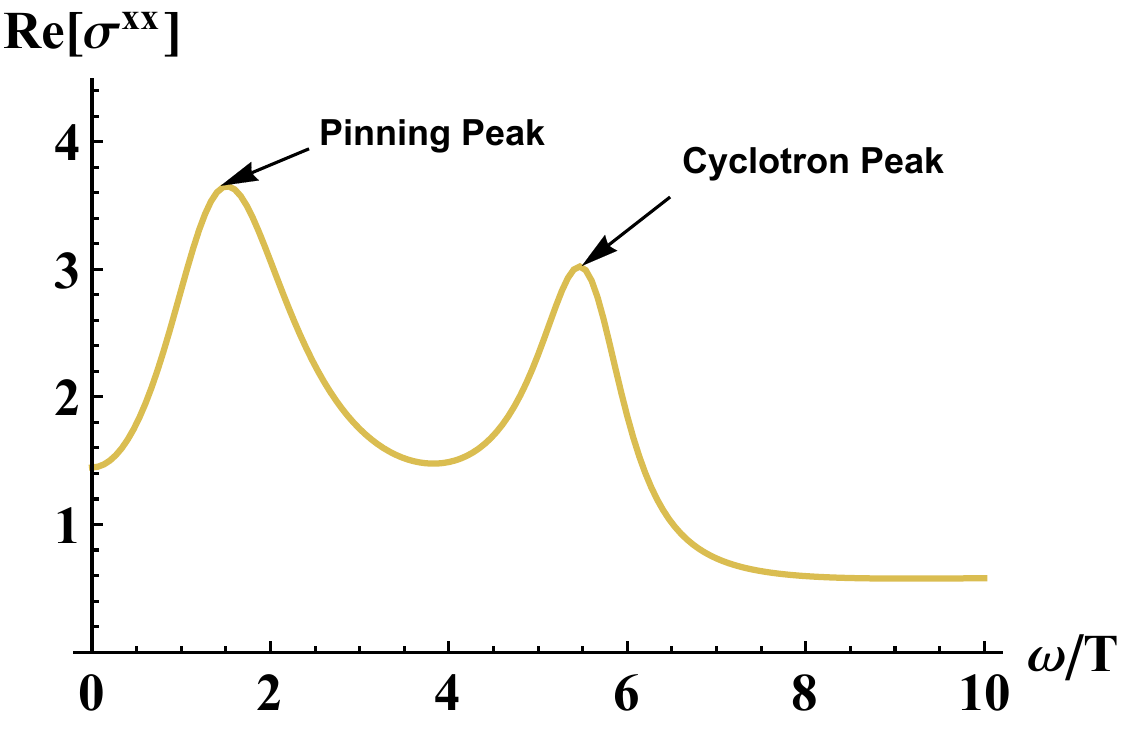} \label{}}
	\subfigure[$H$ evolution for $q_{\chi}=10$]
	{\includegraphics[width=4.8cm]{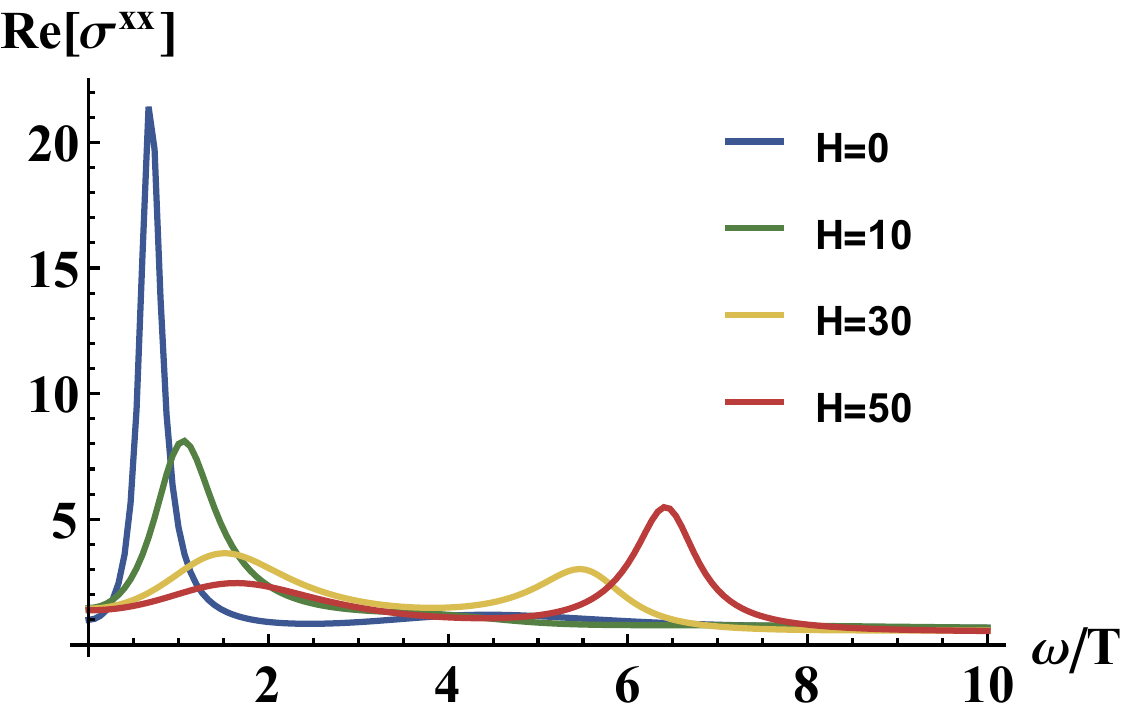} \label{}}
		\subfigure[$q_{\chi}$ evolution for $H=30$ ]
	{\includegraphics[width=4.8cm]{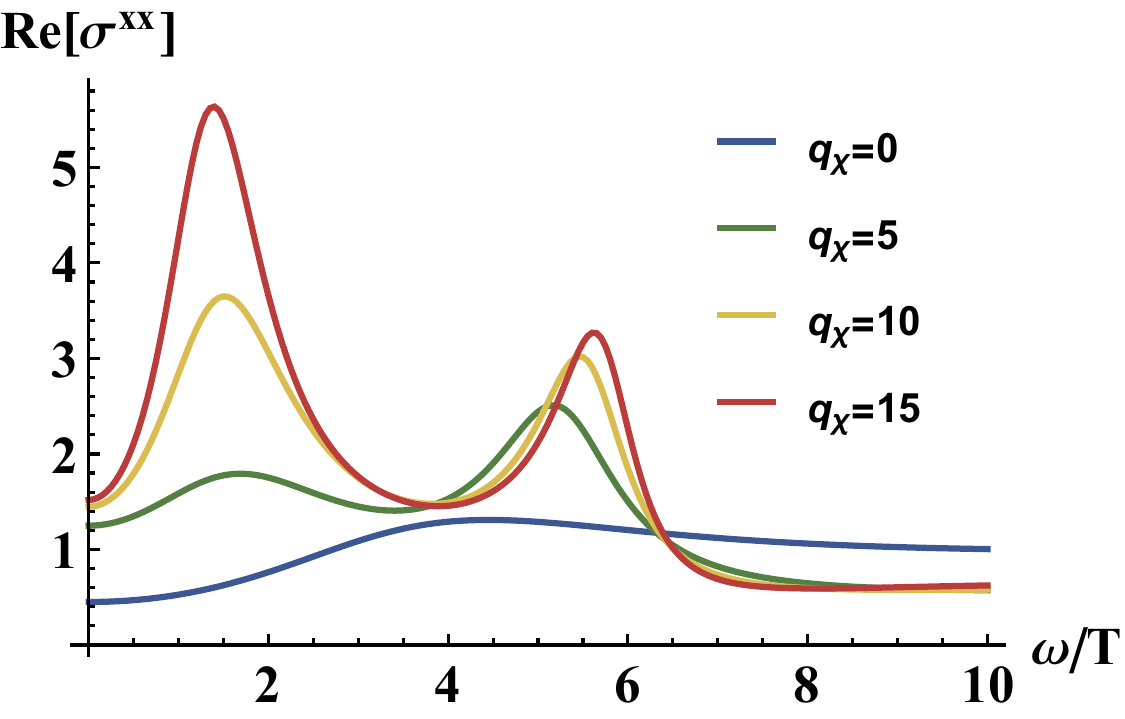} \label{}}
	\caption{pinning peak v.s cyclotron peak : (a) Representative figure for pinning peak and cyclotron peak. $q_{\chi}=10$ and $H=30$.  (b) $H$ evolution with fixed $q_{\chi}=10$. (c) $q_{\chi}$ evolution with fixed $H=30$. 	
	} \label{fig:PvsC}
\end{figure}         
Now we study the effect of magnetic field  on the optical conductivity in the absence of chemical potential.  In our model, $\beta$ plays role of magnetic impurity density  $q_{\chi}$ is its coupling with instanton density. One can ask whether the new peak in previous section, which we called as pinning peak,  can be separated from the cyclotron peak which appears when one  turns on the external magnetic field.  The external magnetic field dependence of the cyclotron pole in the holographic context was  already studied   in \cite{Hartnoll:2007aa,Kim:2015wba} before.

Figure \ref{fig:PvsC} (a)   shows the pinning and cyclotron peak simultaneously. To see that these peaks are different, we study two cases i) change the value of $q_{\chi}$ at fixed $H$, ii) change the external magnetic field at the fixed coupling.  

Figure \ref{fig:PvsC} (b) shows  $H$-dependence of two peaks for fixed $q_{\chi}$.  For $H=0$,  only pinning peak exists as shown in the previous subsection. As we increase the external magnetic field, a new peak appears at higher frequency than that of the pinning peak. The frequency and the height of this peak increases as the external magnetic field is increased. This is  the same  behavior of the cyclotron peak in the dyonic black hole \cite{Hartnoll:2007aa,Kim:2015wba}. Therefore, we can say that the higher frequency peak is a cyclotron peak. Interestingly, as we increase the external magnetic field,  the pinning peak is suppressed. 

Figure \ref{fig:PvsC} (c) shows the $q_{\chi}$ dependence of the two peaks in the presence of the external magnetic field. For $q_\chi=0$ (blue line in the figure), there is a single cyclotron peak as it happens in the dyonic black hole. 
As we increase the value of $q_{\chi}$, the pinning peak is developed at lower frequency region. Simultaneously,  the cyclotron peak gets also shaper. The latter phenomena can be understood as follows; the effect of $q_{\chi}$ gives  anomalous magnetization, a finite magnetization   in the absence of the external magnetic field \cite{Seo:2015pug,Seo:2017yux}.  Therefore, effective magnetic field  increases as we increase $q_{\chi}$ at the fixed  $H$. It makes the cyclotron peak   sharper. Similar behavior happens in very large $q_{\chi}$ without external magnetic field.

\begin{figure}[t!]
\centering
    \subfigure[Pinning pole]
   {\includegraphics[width=5cm]{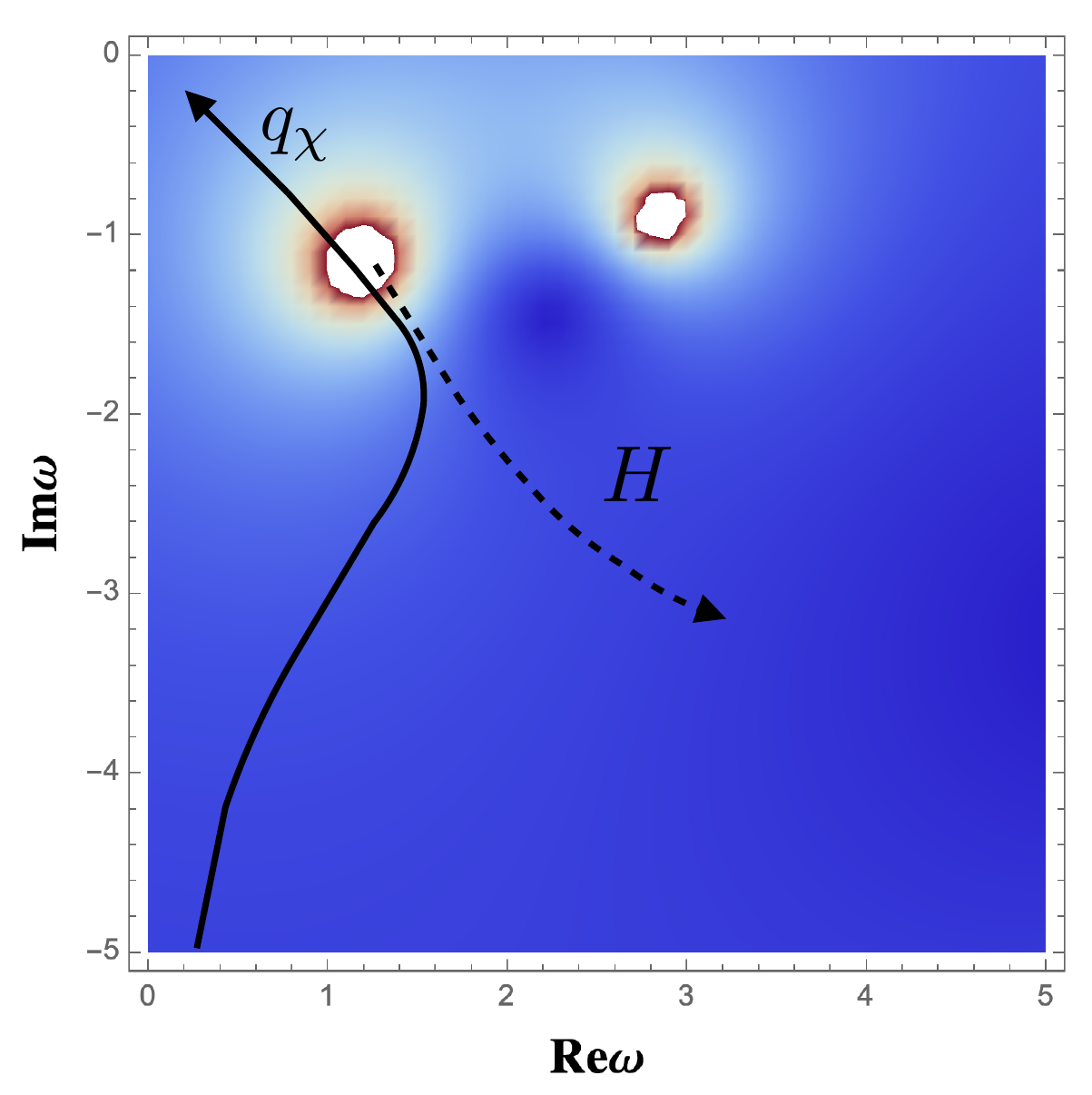} \label{}}
   \hspace{1.5cm}
       \subfigure[Cyclotron pole]
   {\includegraphics[width=5cm]{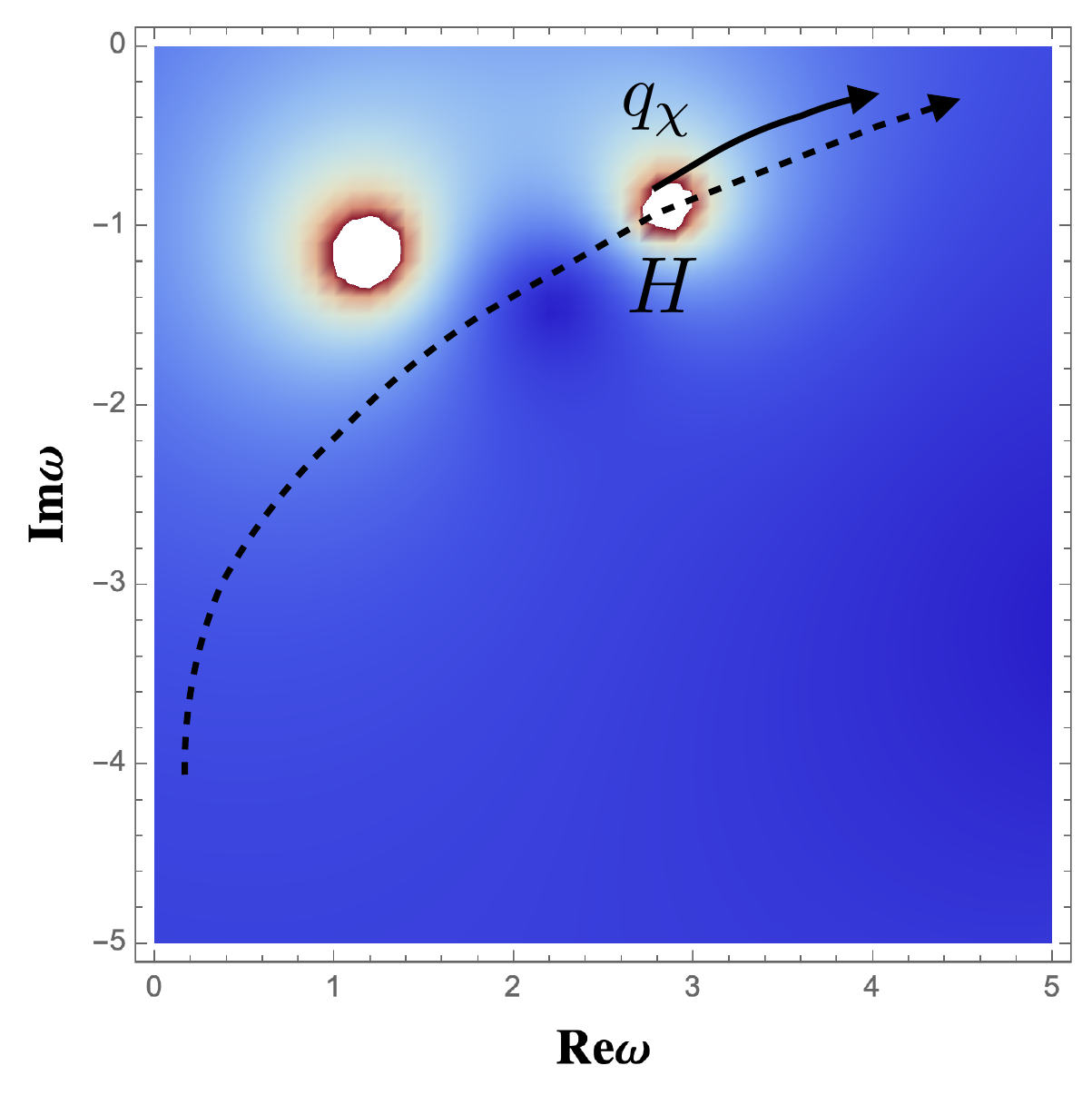} \label{}}
   \caption{ The evolution of the quasi-normal mode as $q_{\chi}$ (solid arrow)  and $H$( dashed arrow ) increases.   (a) For the pinning pole and (b) For the cyclotron pole. In both figures $\beta/T=3$, $q_{\chi}=10$ and $H/T^2=10$. 
           } \label{fig:densityqH}
           \end{figure}
One way to investigate pole structure is calculating $\sigma_{+} \equiv \sigma_{xy} +i\sigma_{xx}$ in complex $\omega$ plane \cite{Hartnoll:2007aa}.
Figure \ref{fig:densityqH} shows density plot of $|\sigma_{+}|$ at $\beta/T=3$, $q_{\chi}=10$ and $H=10$. For given parameters, there are two poles in complex $\omega$ plane.  The pole nearer to the origin corresponds to the  `pinning pole'. The other pole  is the cyclotron pole.  Two arrowed lines indicate the movement of each pole  as we increase $q_{\chi}$ (solid line) or   $H$ (dashed line). 
The movement of the cyclotron peak is relatively simpler. 
It moves away from from the origin and approaches to the real line as we increases either $q_{\chi}$ or $H$. 
 On the other hand,  pinning pole moves in more complicated way under the increase of $q_{\chi}$:  
It approaches to the real line but its real part increases at first but decreases after a critical point.    Therefore,  as $H$ increases, the cyclotron peak is getting sharper,  but the pinning peak is getting   fuzzier. As $q_{\chi}$ increases, both peaks are getting sharper. 

\subsection{ Drude  vs. Pinning peak: $\mu \ne 0$ and $H =0$}
In this subsection, we investigate an effect of $\beta$ and $q_{\chi}$ with finite chemical potential. In the absence of $q_{\chi}$-term and the external magnetic field $H$, the background geometry becomes RN-AdS black hole with momentum relaxation. The optical conductivity in this background is studied in \cite{Kim:2014bza}.  The optical conductivity   changes from the coherent metalic state with      the Drude peak to the incoherent state as the momentum relaxation parameter $\beta$ increases.

In the absence of the external magnetic field, the DC conductivity is independent of $q_{\chi}$ and remains the same as  that  of RN-AdS case \cite{Seo:2017yux} as one can see from 
\begin{align}\label{eq:DC0}
\sigma_{xx}\Big|_{H=0} = 1+\frac{\mu^2}{\beta^2}.
\end{align}
\begin{figure}[b!]
\centering
    \subfigure[$\mu/T=10$, $\beta/T$=5]
   {\includegraphics[width=6cm]{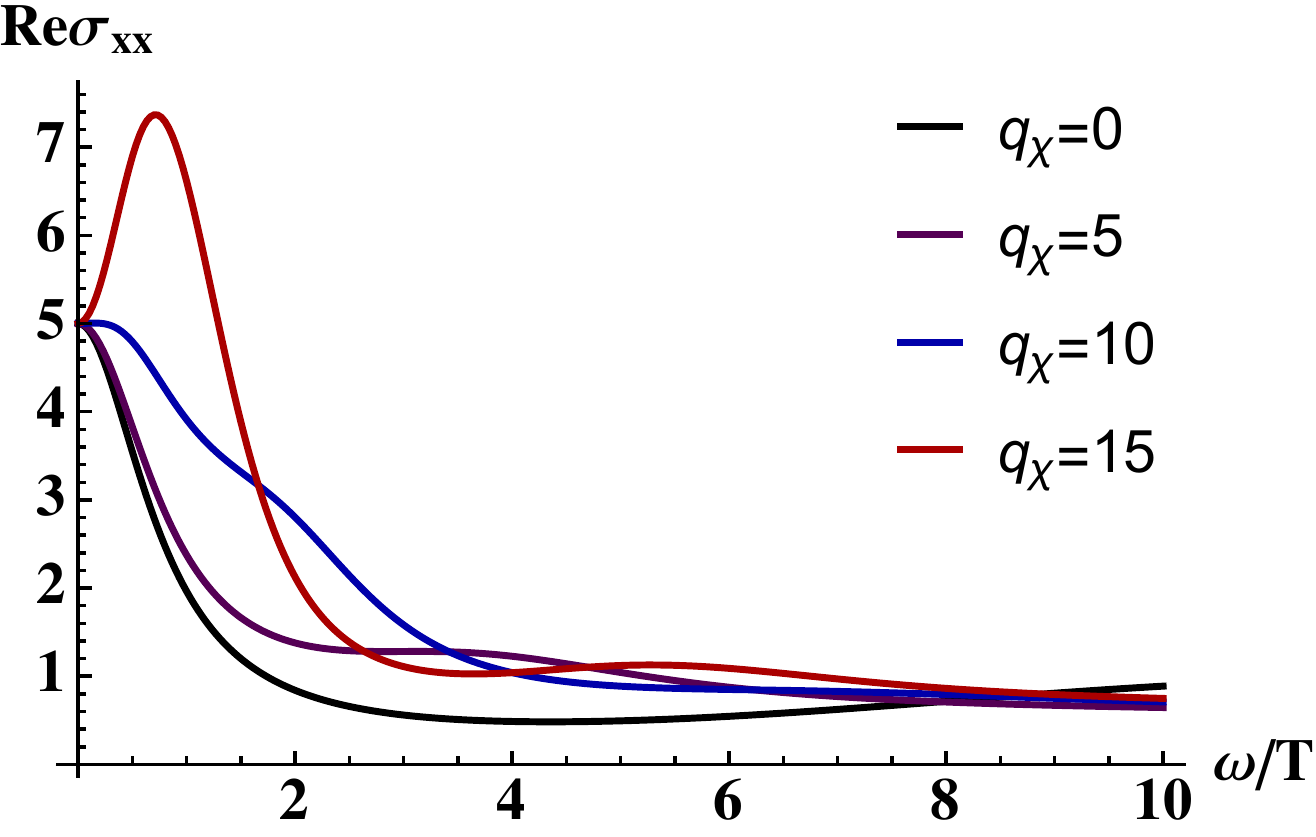} \label{}}
   \hspace{1.5cm}
       \subfigure[$\mu/T=10$, $q_{\chi}=5$]
   {\includegraphics[width=6cm]{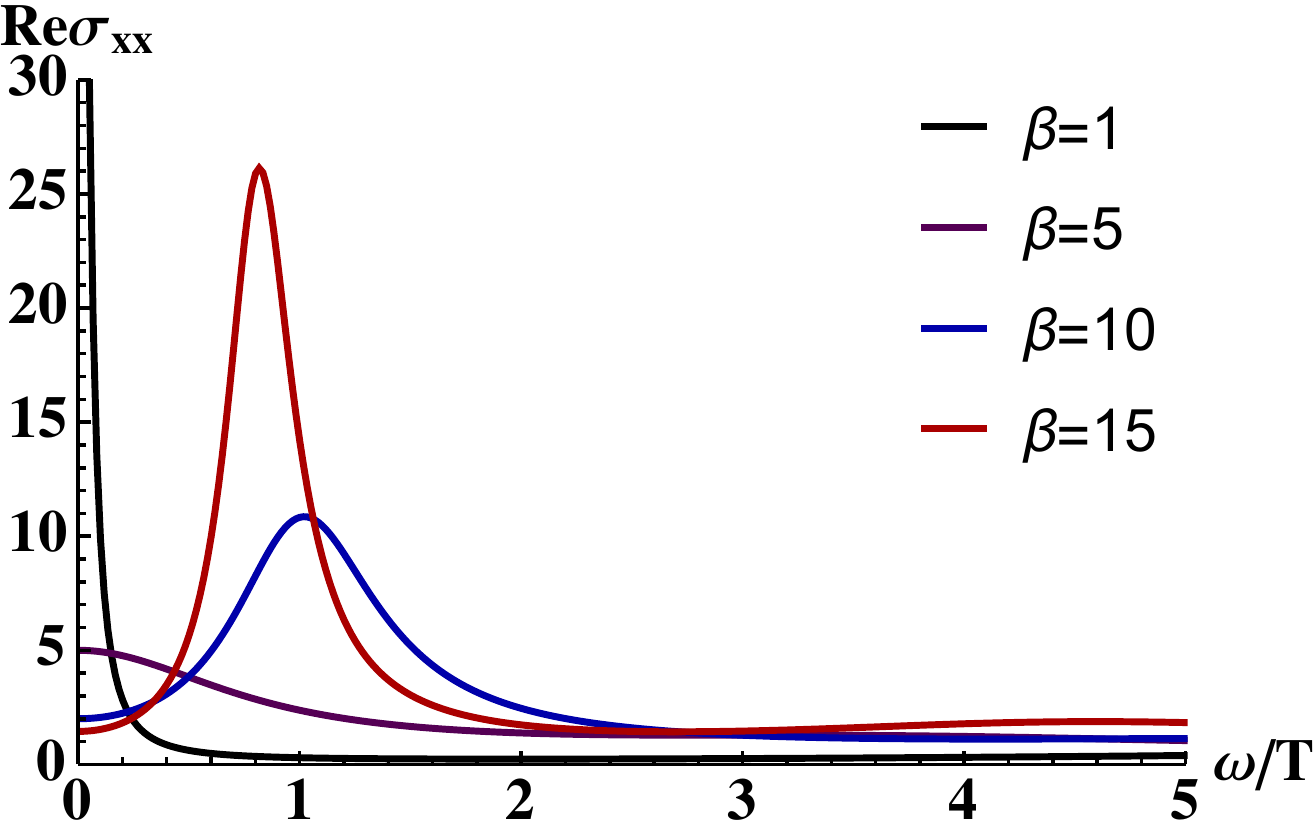} \label{}}
   \caption{ (a) $q_{\chi}$ dependence of optical conductivity for $\mu/T=10$, $\beta/T=5$ and $H/T^2=0$. (b) $\beta$ dependence of optical conductivity for $\mu/T=10$, $q_{\chi}=5$ and $H=0$.
           } \label{fig:CBmu}
           \end{figure}
Figure \ref{fig:CBmu} (a) shows $q_{\chi}$ evolution of the optical conductivity for given chemical potential and $\beta$. In this case, the DC conductivity is fixed by (\ref{eq:DC0}). As $q_{\chi}$ increased, the pinning peak is developed while the Drude peak does not changed. Similarly to the previous sections the pinning frequency decreases as $q_{\chi}$ increases and finally the Drude peak is observed. 
Figure \ref{fig:CBmu} (b) is $\beta$ dependence of the optical conductivity with fixed chemical potential and $q_{\chi}$. Here,  the DC conductivity is suppressed by (\ref{eq:DC0}) and the pinning peak is also developed. The behavior of the pinning peak is similar to the evolution of $q_{\chi}$ because the $q_{\chi}$ interaction term in the action (\ref{eq:action}) also contains scalar term $(\partial \chi)^2$ which is proportional to $\beta^2$. Therefore, increasing $\beta$ effect should be the same as increasing $q_{\chi}$. 

   \begin{figure}[]
\centering
    \subfigure[$\mu/T=10$, $\beta/T$=5]
   {\includegraphics[width=6cm]{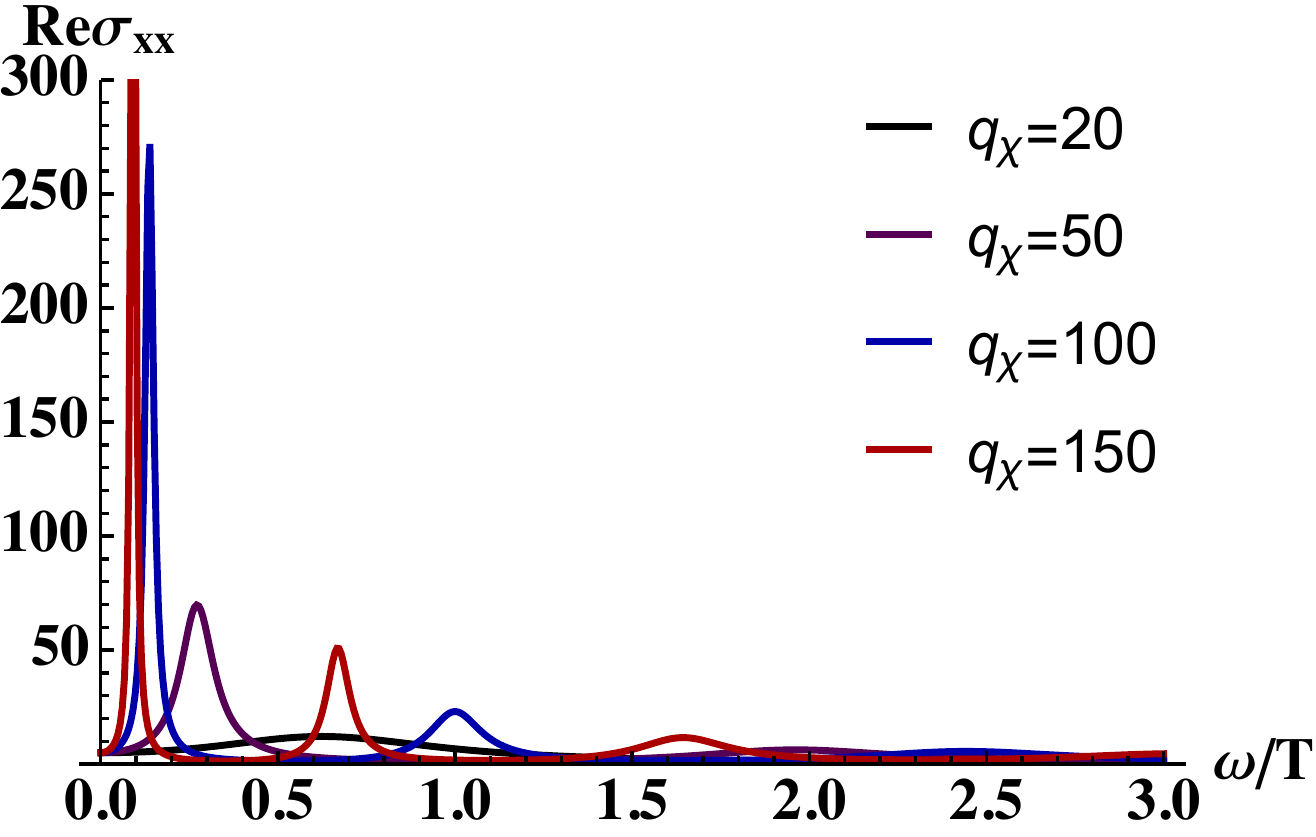} \label{}}
   \hspace{1.5cm}
       \subfigure[$\mu/T=10$, $q_{\chi}=5$]
   {\includegraphics[width=6cm]{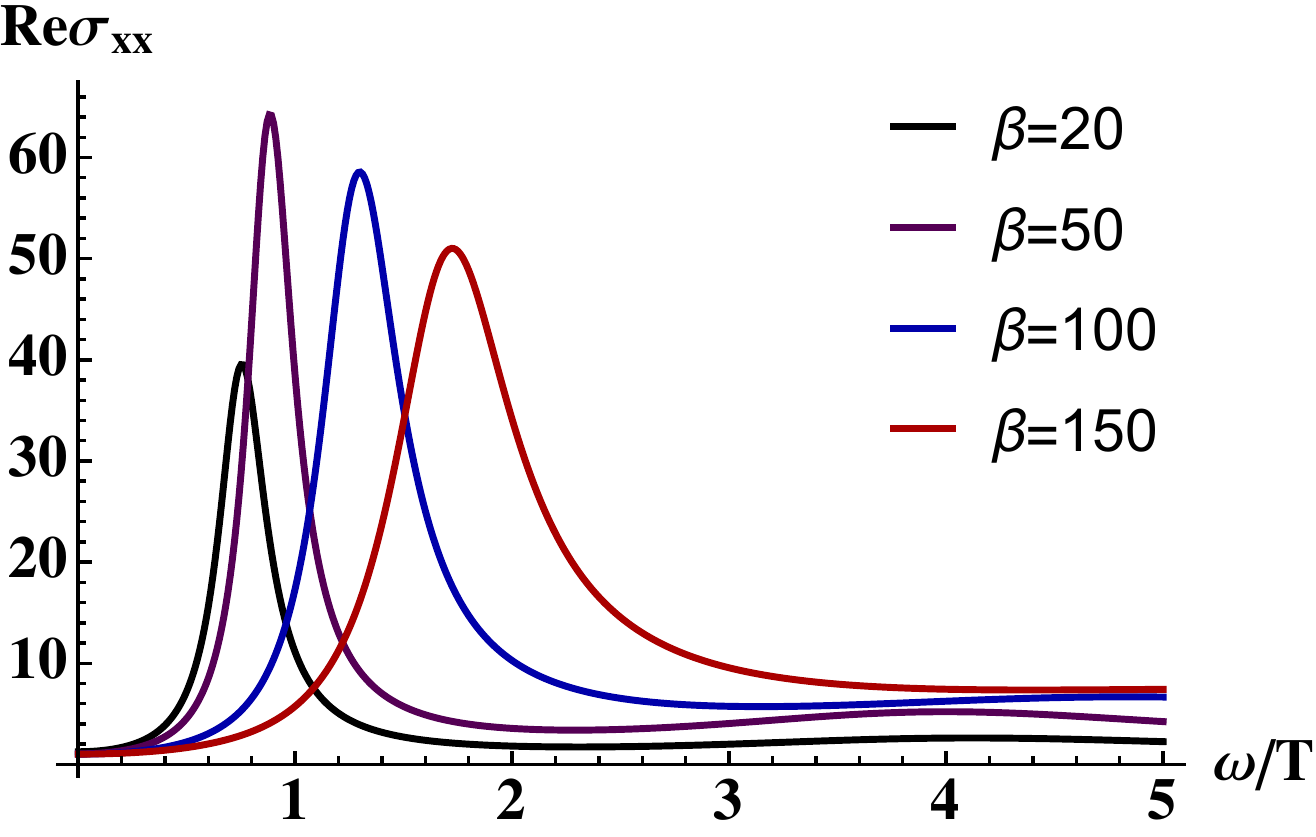} \label{}}
   \caption{ (a)  $q_{\chi}$ evolution of optical conductivity for
   large $q_{\chi}$.  $\mu/T=10$, $\beta/T=5$ and $H=0$. Notice that number of peaks increases as $q_{\chi}$ increases. (b)  $\beta$ evolution of optical conductivity for large $\beta$. Notice that peak frequency increases as function of $\beta$. $\mu/T=10$, $q_{\chi}=5$ and $H=0$.
           } \label{fig:CBLarge}
           \end{figure}    

One should notice, however, that  the large $\beta$ behavior is different to large $q_{\chi}$. See Figure \ref{fig:CBLarge}.        
For a large value of $q_{\chi}$, the peak frequency of the pinning peak is decreasing and the height is increasing. At the same time, another cyclotron peak appear at high frequency region as shown in \ref{fig:CBLarge} (a). The appearance of new cyclotron peak   comes from the finite magnetization effect with finite $q_{\chi}$. This magnetization gives effective magnetic field and hence a new cyclotron peak appear. On the other hand, when $\beta$ is increased, the pinning peak becomes sharper first(Figure \ref{fig:CBmu} (b)). But if we increase $\beta$ more, the pinning peak is suppressed as Figure \ref{fig:CBLarge} (b). This behavior can be understood as follows;  $\beta$ plays a role of momentum relaxation parameter as well as a magnetic impurity via  coupling  term in the action. The pinning effect tends to make the peak sharp while the momentum relaxation effect tends to suppress the peak. Therefore, there is a competition between the pinning effect and the impurity effect. When $\beta$ is small, the pinning effect is stronger than the impurity effect and hence increasing $\beta$ is similar to increasing $q_{\chi}$. For a  larger $\beta$,  its effect is to suppress the pinning pole: the peak frequency goes high and the peak height is decreased. See figure \ref{fig:CBLarge}(b).

\begin{figure}[t!]
\centering
    \subfigure[$q_{\chi}$ evolution]
   {\includegraphics[width=5cm]{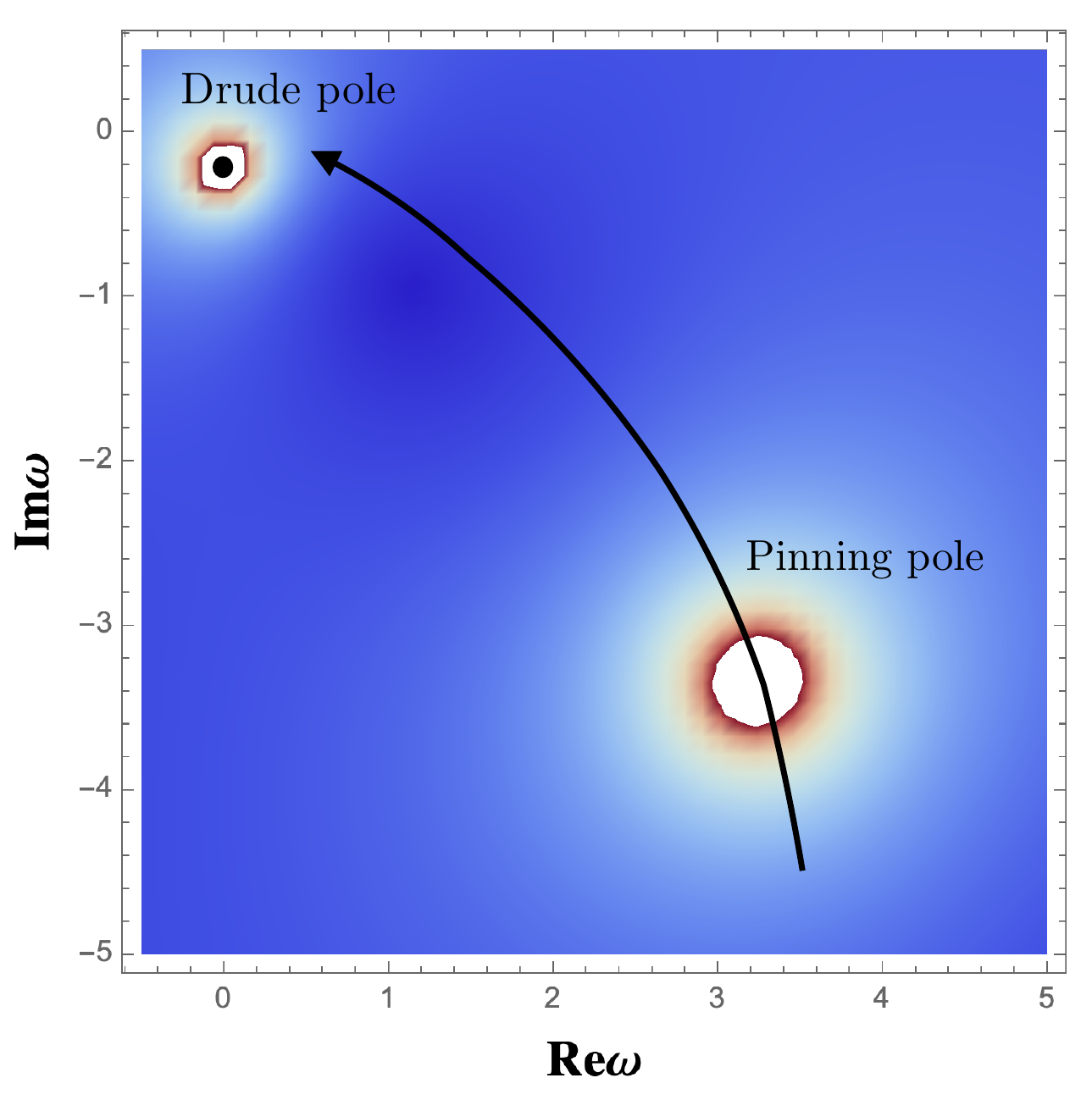} \label{}}
   \hspace{1.5cm}
       \subfigure[$\beta$ evolution]
   {\includegraphics[width=5cm]{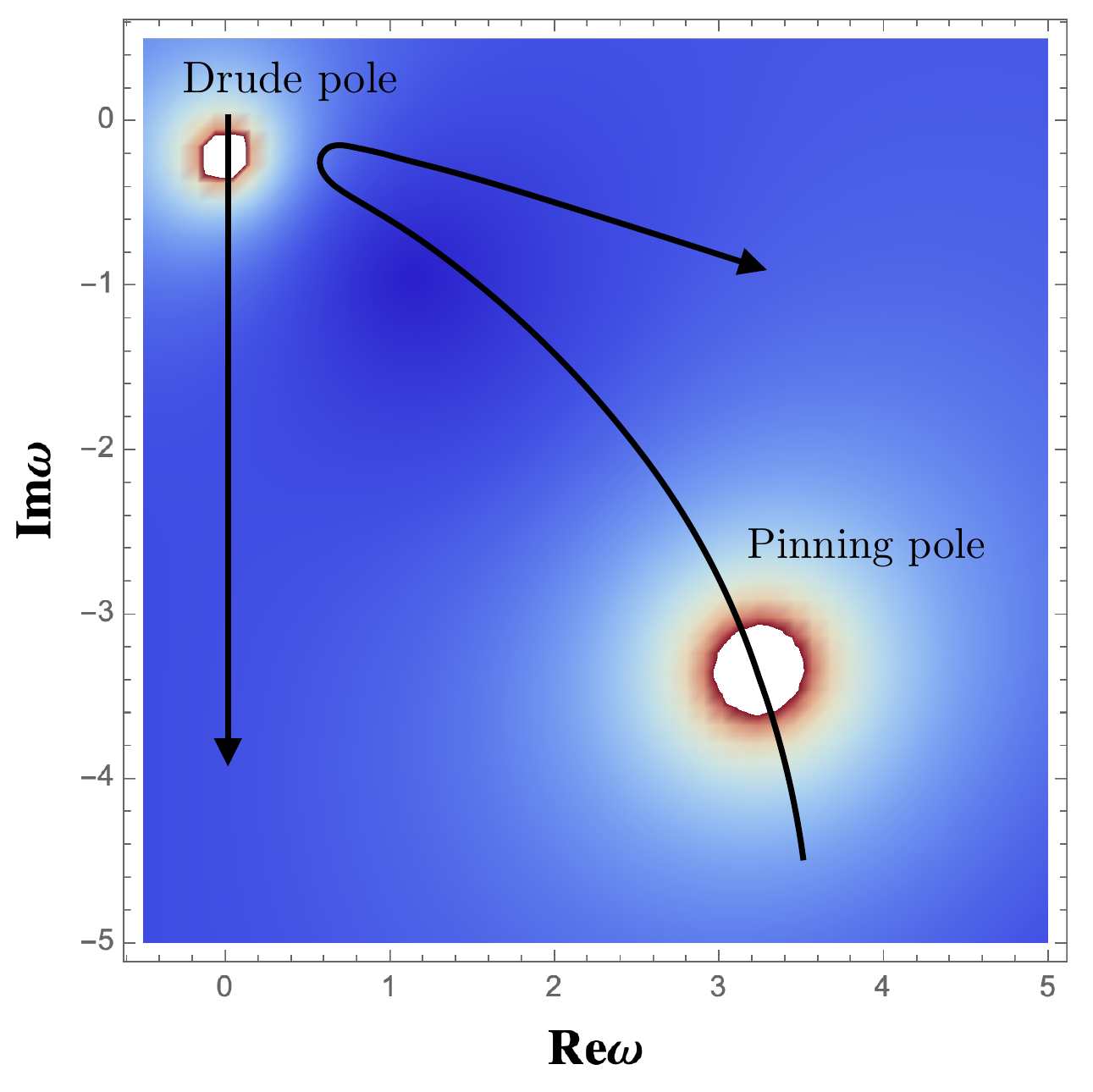} \label{}}
   \caption{ (a) $q_{\chi}$ evolution of the Drude and the pinning pole. (b) $\beta$ evolution of the Drude and the pinning pole. 
    The density plot is drawn with $\mu/T=5$, $\beta/T =2$, $q_{\chi}=5$ and $H=0$.
           } \label{fig:densityqb}
           \end{figure}

Figure \ref{fig:densityqb} is a density plot of $|\sigma_{+}|$ at $\mu/T=5$, $\beta/T=1$ and $q_{\chi}=10$. 
When $\mu^2 \gg \beta^2$, there is a sharp Drude peak in the optical conductivity. In complex $\omega$ plane, we call `Drude pole' if the  pole is near  origin.
The pinning pole is at finite complex $\omega$. Figure \ref{fig:densityqb} (a) shows $q_{\chi}$ evolution of the Drude and the pinning pole. As $q_{\chi}$ increases the pinning pole approaches to the origin of the complex $\omega$ plane while the Drude pole does not move. In numerical calculation, the residue  of the pinning pole is much larger than the Drude pole. 
Therefore, the Drude pole is almost  overlapping with the pinning pole at large value of $q_{\chi}$ which is consistent with Figure~\ref{fig:CBmu}(a).

$\beta$ evolution of the Drude and the pinning pole is drawn in Figure \ref{fig:densityqb} (b).  In the figure, the solid arrowed line denotes the direction of each pole as $\beta$ is increased.  As $\beta$ is increased, the Drude pole move down along the imaginary $\omega$ axis which corresponds to the suppression of the real part of the conductivity. 
The $\beta$-evolution   of the pinning pole is more non-trivial: at first, the pinning pole approaches to the origin just like the case of the $q_{\chi}$ evolution. But if we increase $\beta$ more, the pinning pole changes its direction to positive $\omega$ and the image on the real $\omega$ axis becomes Figure \ref{fig:CBLarge} (b).

One remark is that  there is a phase transition in $\beta$ evolution: for small $\beta$ the Drude peak is sharp therefore it is metal but for large values of $\beta$ the Drude pole is moved below so that it is insulator or bad metal at best.  See Figure \ref{fig:CBmu}(b).
Therefore, describing the coupling dependence of the Drude peak is a good way to describe the Metal-insulator transition in the model.

  \subsection{Origin of pinning pole}
Now it is time to discuss what the origin and nature of the new pinning peak are.
Since we   look at the dynamics of current operators  $J_{\mu}$ at the boundary, we consider their bulk dual field $A_{\mu}$. We first  observe from the eq.(\ref{max1})  and  eq.(\ref{em1}) that for the cyclotron peak $a_x$ is related to  $a_y$ through the $h_{tx}$ to create a vortex, while for the  
pinning peak,   $a_{x},a_{y}$  are connected by the $q_{chi} (\beta^2) $ term directly. 
Such difference is what makes the   behaviors of the two poles different.  In other words, the pinning pole is nothing but the cyclotron peak by the anomalous magnetic moment induced by the interaction term. 
In fact the term 
\be
 \epsilon_{ij}\left(\frac{2i\theta\omega}{r^3U}a_j+\frac{iH\omega}{U^2}h_{tj}\right)  
\ee
  in the eq.(\ref{max1})  shows the parallelism of 
$\theta a_{y}$ and $Hh_{ty}$ regarding their coupling to   $a_{x}$. 

However, one can ask that  if the pinning pole is nothing but the 
cyclotron vortex due to the anomalous magnetization, how two poles can be created.  In fact, if   we   assume that only the   magnetic induction  $B$, which is the sum of magnetization $M$ and the current-induced magnetic field $H$,  can couple to the charged particle, then  there can not be two separate cyclotron poles. 
However, such splitting $B=M+H$   is possible only when the interaction is the minimal one $A_{\mu}J^{\mu}$  and  the source  $J^{\mu}$  can be split into charge's motional current   and magnetization current:  $J^{\mu}=J^{\mu}_{c}+J^{\mu}_{m}$ with  $J^{\mu}_{m}=\partial_{\nu}M^{\nu\mu}$ for some antisymmetric $M^{\nu\mu}$, so that the equation of motion
can be rewritten as  
\be 
\partial_{\nu} (F^{\mu\nu}-M^{\mu\nu}) = J^{\mu}_{c}.
\ee
Taking the space-space components of $F^{\mu\nu}$ and $M^{\mu\nu}$ as the $B$ and $M$, we get the   splitting $B-M=H$. We want to answer why two poles 
should exist. 

The basic point is that such canonical splitting is impossible when we add  a non-minimal interaction  like  Chern-Simon term  or   $q_{\chi}$ type coupling we used.  In that case, $M$ and $B$ can develop  independent cyclotron peaks.  
In fact, for all recent studies on holographic pinning effect, the interactions considered  were in the category of non-minimal interaction of gauge fields,  where we expect the  presence of pinning effects. 

Our observation explains the origin of pinning peak not only for our case but also for all other case as well. For example,  if we  add Chern-Simon type interaction to induce the metal insulator transition\cite{Donos:2013gda}, such appearance of additional pole, which we call as pinning peak, is generic and unavoidable as it was observed in \cite{Jokela:2014dba,Ling:2014saa,Donos:2014oha,Cai_2017} with a few different types of interactions. 
Such shifted peak also arises when one calculate  the holographic transport using the higher derivative terms, see \cite{Myers:2010pk}, which is another supporting evidence of our claim. 

Finally we comment that although the physics of pinning peak in our case  is described as the cyclotron peak by the anomalous magnetization($M$ for $H=0$),  for some other cases, it might  be described as the physics of anomalous polarization by taking a time-space component of  $F^{\mu\nu}, M^{\mu\nu}$  so that $E+P=D$.

\section{Discussion}
In this paper, 
 we proposed   the role of a bulk interaction term as a creator of new spectrum. 
 Since we are using a Lagrangian which is similar to the one used to discuss the metal-insulator transition  or pinning effect, we want to comment  on the relation of our result with them. 
 
 We described   the appearance of a sharp peak in holography, but it is an example of spectral change of many body theory.   For any theory respecting unitarity, the changes of the spectrum distribution occur under the constraint of a conductivity sum rule so the  spectral change  is just rearrangement of the degrees of freedom.  
The conductor/insulator distinction depends on whether or not  the spectral peak  goes through $\omega=0$,  regardless of    strength of the correlation.
 
The pinning effect  requests the appearance of the new peak at $\omega\neq 0$ but it   requests disappearance of the Drude peak too,  so it can be simply described as a shift of the Drude peak to somewhere else. 
 For the metal to insulator transition (MIT),  
it should  come through the gap creation in the optical conductivity. Conversely, the insulator to metal  transition requests the creation of the  peak at $\omega=0$ of optical conductivity. 
This is common whatever the origin of that MIT is: 
all microscopic mechanism  including  interaction induced (Mott insulator), impurity induced (Anderson insulator) and structural change induced (Band insulator) are just a way of creating/deleting the complex pole near $\omega=0$ and   eventually   should be described  in terms of the spectral change. 

The idea  of changing the spectrum 
in terms of holographic bulk  interaction can be useful for all above mentioned microscopic mechanism,  because in holography, the bulk locality  is consequence of boundary non-locality~\cite{Hamilton:2006az}. The bulk local interaction may  be able to describe the global structure change or randomly distributed disorder as well as the Plankian dissipation in strange metal  that apparently require the non-local interaction.   
 
The bulk interaction changes the character of the system and we expect that it would be  interesting to utilize this to characterize a system and to match it to a realistic system in the future.   In this regards, it would be interesting to examine  all the possible lowest interaction terms of fermions and gauge fields and check whether each interaction term really generates new branch of quasi particle spectrums.

\acknowledgments
 We would like to thank M. Baggioli for interesting comments on our first version. 
This  work is supported by Mid-career Researcher Program through the National Research Foundation of Korea grant No. NRF-2016R1A2B3007687.  YS is  supported  by Basic Science Research Program through NRF grant No. NRF-2016R1D1A1B03931443. The work of K.-Y. Kim was supported by Basic Science Research Program through the National Research Foundation of Korea(NRF) funded by the Ministry of Science, ICT $\&$ Future Planning(NRF- 2017R1A2B4004810) and GIST Research Institute(GRI) grant funded by the GIST in 2019. 

\bibliographystyle{JHEP}
\bibliography{Refs_AC}

\providecommand{\href}[2]{#2}\begingroup\raggedright\begin{thebibliography}{10}

\bibitem{Iqbal:2009fd}
N.~Iqbal and H.~Liu, \emph{{Real-time response in AdS/CFT with application to
  spinors}}, \href{http://dx.doi.org/10.1002/prop.200900057}{\emph{Fortsch.
  Phys.} {\bf 57} (2009) 367--384}, [\href{http://arxiv.org/abs/0903.2596}{{\tt
  0903.2596}}].

\bibitem{Edalati:2010ww}
M.~Edalati, R.~G. Leigh and P.~W. Phillips, \emph{{Dynamically Generated Mott
  Gap from Holography}},
  \href{http://dx.doi.org/10.1103/PhysRevLett.106.091602}{\emph{Phys. Rev.
  Lett.} {\bf 106} (2011) 091602}, [\href{http://arxiv.org/abs/1010.3238}{{\tt
  1010.3238}}].

\bibitem{Seo:2018hrc}
Y.~Seo, G.~Song, Y.-H. Qi and S.-J. Sin, \emph{{Mott transition with
  Holographic Spectral function}},  \href{http://arxiv.org/abs/1803.01864}{{\tt
  1803.01864}}.

\bibitem{Seo:2015pug}
Y.~Seo, K.-Y. Kim, K.~K. Kim and S.-J. Sin, \emph{{Character of matter in
  holography: Spin--orbit interaction}},
  \href{http://dx.doi.org/10.1016/j.physletb.2016.05.059}{\emph{Phys. Lett.}
  {\bf B759} (2016) 104--109}, [\href{http://arxiv.org/abs/1512.08916}{{\tt
  1512.08916}}].

\bibitem{Seo:2017yux}
Y.~Seo, G.~Song, C.~Park and S.-J. Sin, \emph{{Small Fermi Surfaces and Strong
  Correlation Effects in Dirac Materials with Holography}},
  \href{http://dx.doi.org/10.1007/JHEP10(2017)204}{\emph{JHEP} {\bf 10} (2017)
  204}, [\href{http://arxiv.org/abs/1708.02257}{{\tt 1708.02257}}].

\bibitem{Nakamura:2009tf}
S.~Nakamura, H.~Ooguri and C.-S. Park, \emph{{Gravity Dual of Spatially
  Modulated Phase}},
  \href{http://dx.doi.org/10.1103/PhysRevD.81.044018}{\emph{Phys. Rev.} {\bf
  D81} (2010) 044018}, [\href{http://arxiv.org/abs/0911.0679}{{\tt
  0911.0679}}].

\bibitem{Donos:2013gda}
A.~Donos and J.~P. Gauntlett, \emph{{Holographic charge density waves}},
  \href{http://dx.doi.org/10.1103/PhysRevD.87.126008}{\emph{Phys. Rev.} {\bf
  D87} (2013) 126008}, [\href{http://arxiv.org/abs/1303.4398}{{\tt
  1303.4398}}].

\bibitem{Erdmenger:2013zaa}
J.~Erdmenger, X.-H. Ge and D.-W. Pang, \emph{{Striped phases in the holographic
  insulator/superconductor transition}},
  \href{http://dx.doi.org/10.1007/JHEP11(2013)027}{\emph{JHEP} {\bf 11} (2013)
  027}, [\href{http://arxiv.org/abs/1307.4609}{{\tt 1307.4609}}].

\bibitem{Jokela:2014dba}
N.~Jokela, M.~Jarvinen and M.~Lippert, \emph{{Gravity dual of spin and charge
  density waves}}, \href{http://dx.doi.org/10.1007/JHEP12(2014)083}{\emph{JHEP}
  {\bf 12} (2014) 083}, [\href{http://arxiv.org/abs/1408.1397}{{\tt
  1408.1397}}].

\bibitem{Ling:2014saa}
Y.~Ling, C.~Niu, J.~Wu, Z.~Xian and H.-b. Zhang, \emph{{Metal-insulator
  Transition by Holographic Charge Density Waves}},
  \href{http://dx.doi.org/10.1103/PhysRevLett.113.091602}{\emph{Phys. Rev.
  Lett.} {\bf 113} (2014) 091602}, [\href{http://arxiv.org/abs/1404.0777}{{\tt
  1404.0777}}].

\bibitem{Donos:2014oha}
A.~Donos, B.~Goutraux and E.~Kiritsis, \emph{{Holographic Metals and Insulators
  with Helical Symmetry}},
  \href{http://dx.doi.org/10.1007/JHEP09(2014)038}{\emph{JHEP} {\bf 09} (2014)
  038}, [\href{http://arxiv.org/abs/1406.6351}{{\tt 1406.6351}}].

\bibitem{Cai_2017}
R.-G. Cai, L.~Li, Y.-Q. Wang and J.~Zaanen, \emph{Intertwined order and
  holography: The case of parity breaking pair density waves},
  \href{http://dx.doi.org/10.1103/physrevlett.119.181601}{\emph{Physical Review
  Letters} {\bf 119} (Nov, 2017) }.

\bibitem{Donos:2012js}
A.~Donos and S.~A. Hartnoll, \emph{{Interaction-driven localization in
  holography}}, \href{http://dx.doi.org/10.1038/nphys2701}{\emph{Nature Phys.}
  {\bf 9} (2013) 649--655}, [\href{http://arxiv.org/abs/1212.2998}{{\tt
  1212.2998}}].

\bibitem{Jokela_2017}
N.~Jokela, M.~J{\"a}rvinen and M.~Lippert, \emph{Pinning of holographic sliding
  stripes}, \href{http://dx.doi.org/10.1103/physrevd.96.106017}{\emph{Physical
  Review D} {\bf 96} (Nov, 2017) }.

\bibitem{Andrade_2018}
T.~Andrade, M.~Baggioli, A.~Krikun and N.~Poovuttikul, \emph{Pinning of
  longitudinal phonons in holographic spontaneous helices},
  \href{http://dx.doi.org/10.1007/jhep02(2018)085}{\emph{Journal of High Energy
  Physics} {\bf 2018} (Feb, 2018) }.

\bibitem{Donos:2019hpp}
A.~Donos, D.~Martin, C.~Pantelidou and V.~Ziogas, \emph{{Incoherent
  hydrodynamics and density waves}},
  \href{http://arxiv.org/abs/1906.03132}{{\tt 1906.03132}}.

\bibitem{SciPostPhys.3.3.025}
L.~V. Delacr{\'e}taz, B.~Gout{\'e}raux, S.~A. Hartnoll and A.~Karlsson,
  \emph{{Bad Metals from Fluctuating Density Waves}},
  \href{http://dx.doi.org/10.21468/SciPostPhys.3.3.025}{\emph{SciPost Phys.}
  {\bf 3} (2017) 025}.

\bibitem{PhysRevB.17.535}
H.~Fukuyama and P.~A. Lee, \emph{Dynamics of the charge-density wave. i.
  impurity pinning in a single chain},
  \href{http://dx.doi.org/10.1103/PhysRevB.17.535}{\emph{Phys. Rev. B} {\bf 17}
  (Jan, 1978) 535--541}.

\bibitem{PhysRevLett.45.935}
G.~Gr\"uner, L.~C. Tippie, J.~Sanny, W.~G. Clark and N.~P. Ong,
  \emph{Frequency-dependent conductivity in nb${\mathrm{se}}_{3}$},
  \href{http://dx.doi.org/10.1103/PhysRevLett.45.935}{\emph{Phys. Rev. Lett.}
  {\bf 45} (Sep, 1980) 935--938}.

\bibitem{Fukuyama:1978ab}
H.~Fukuyama, \emph{Pinning and conductivity of two-dimensional charge-density
  waves in magnetic fields},
  \href{http://dx.doi.org/10.1103/PhysRevB.18.6245}{\emph{Physical Review B}
  {\bf 18} (1978) 6245--6252}.

\bibitem{Kim:2014bza}
K.-Y. Kim, K.~K. Kim, Y.~Seo and S.-J. Sin, \emph{{Coherent/incoherent metal
  transition in a holographic model}},
  \href{http://dx.doi.org/10.1007/JHEP12(2014)170}{\emph{JHEP} {\bf 12} (2014)
  170}, [\href{http://arxiv.org/abs/1409.8346}{{\tt 1409.8346}}].

\bibitem{Baggioli_2015}
M.~Baggioli and O.~Pujol{\`a}s, \emph{Electron-phonon interactions,
  metal-insulator transitions, and holographic massive gravity},
  \href{http://dx.doi.org/10.1103/physrevlett.114.251602}{\emph{Physical Review
  Letters} {\bf 114} (Jun, 2015) }.

\bibitem{Alberte_2018}
L.~Alberte, M.~Ammon, M.~Baggioli, A.~Jim{\'e}nez and O.~Pujol{\`a}s,
  \emph{Black hole elasticity and gapped transverse phonons in holography},
  \href{http://dx.doi.org/10.1007/jhep01(2018)129}{\emph{Journal of High Energy
  Physics} {\bf 2018} (Jan, 2018) }.

\bibitem{Ammon:2019wci}
M.~Ammon, M.~Baggioli and A.~Jimenez-Alba, \emph{{A Unified Description of
  Translational Symmetry Breaking in Holography}},
  \href{http://arxiv.org/abs/1904.05785}{{\tt 1904.05785}}.

\bibitem{Li_2019}
W.-J. Li and J.-P. Wu, \emph{A simple holographic model for spontaneous
  breaking of translational symmetry},
  \href{http://dx.doi.org/10.1140/epjc/s10052-019-6761-0}{\emph{The European
  Physical Journal C} {\bf 79} (Mar, 2019) }.

\bibitem{Amoretti:2018tzw}
A.~Amoretti, D.~Aren, B.~Goutraux and D.~Musso, \emph{{A holographic strange
  metal with slowly fluctuating translational order}},
  \href{http://arxiv.org/abs/1812.08118}{{\tt 1812.08118}}.

\bibitem{Donos_2019}
A.~Donos and C.~Pantelidou, \emph{Holographic transport and density waves},
  \href{http://dx.doi.org/10.1007/jhep05(2019)079}{\emph{Journal of High Energy
  Physics} {\bf 2019} (May, 2019) }.

\bibitem{Hartnoll:2007aa}
S.~A. Hartnoll and C.~P. Herzog, \emph{Ohm's law at strong coupling: S duality
  and the cyclotron resonance}, {\emph{Phys.Rev.D} {\bf 76} (2007) 106012},
  [\href{http://arxiv.org/abs/0706.3228}{{\tt 0706.3228}}].

\bibitem{Kim:2015wba}
K.-Y. Kim, K.~K. Kim, Y.~Seo and S.-J. Sin, \emph{{Thermoelectric
  Conductivities at Finite Magnetic Field and the Nernst Effect}},
  \href{http://dx.doi.org/10.1007/JHEP07(2015)027}{\emph{JHEP} {\bf 07} (2015)
  027}, [\href{http://arxiv.org/abs/1502.05386}{{\tt 1502.05386}}].

\bibitem{Myers:2010pk}
R.~C. Myers, S.~Sachdev and A.~Singh, \emph{{Holographic Quantum Critical
  Transport without Self-Duality}},
  \href{http://dx.doi.org/10.1103/PhysRevD.83.066017}{\emph{Phys. Rev.} {\bf
  D83} (2011) 066017}, [\href{http://arxiv.org/abs/1010.0443}{{\tt
  1010.0443}}].

\bibitem{Hamilton:2006az}
A.~Hamilton, D.~N. Kabat, G.~Lifschytz and D.~A. Lowe, \emph{{Holographic
  representation of local bulk operators}},
  \href{http://dx.doi.org/10.1103/PhysRevD.74.066009}{\emph{Phys. Rev.} {\bf
  D74} (2006) 066009}, [\href{http://arxiv.org/abs/hep-th/0606141}{{\tt
  hep-th/0606141}}].

\end{thebibliography}\endgroup


\end{document}